\documentclass[3p]{elsarticle}
\journal{Physica A: Statistical Mechanics and its Applications}

\usepackage{amsmath,amssymb,amsthm}
\usepackage{graphicx}
\usepackage{float}
\usepackage{xcolor}
\usepackage{booktabs}
\usepackage{algorithm}
\usepackage{algpseudocode}
\usepackage{microtype}
\usepackage{placeins}
\usepackage{rotating}
\usepackage{hyperref}
\usepackage{orcidlink}

\newtheorem{remark}{Remark}

\newcommand{\bk}{B_k}
\newcommand{\lk}{\lambda_k}
\newcommand{\xtk}{x_t^{(k)}}
\newcommand{\vk}{\mathbf{v}_k}
\newcommand{\et}{\mathbf{e}_t}
\newcommand{\ebar}{\bar{\mathbf{e}}}
\newcommand{\Sig}{\boldsymbol{\Sigma}}
\newcommand{\R}{\mathbb{R}}
\newcommand{\meantau}{\langle\tau\rangle}
\newcommand{\sigmatau}{\sigma_\tau}
\newcommand{\CV}{\mathrm{CV}}

\usepackage{totcount}
\newtotcounter{todocount}

\begin{document}
\begin{frontmatter}

\title{Semantic burstiness across principal directions of textual
variation}

\author[unibo]{Mirko Degli Esposti\corref{cor1}\,\orcidlink{0000-0003-0316-3449}}
\cortext[cor1]{Corresponding author}
\ead{mirko.degliesposti@unibo.it}
\author[ou]{Marcelo A. Montemurro\,\orcidlink{0000-0003-0893-8530}}

\affiliation[unibo]{organization={Department of Physics and Astronomy,
  University of Bologna}, city={Bologna}, country={Italy}}
\affiliation[ou]{organization={School of Mathematics and Statistics,
  The Open University}, addressline={Walton Hall}, city={Milton Keynes},
  country={United Kingdom}}

\begin{abstract}
Long-range correlations in written texts extend over scales far beyond
local syntax, but their relation to semantic organisation remains
incompletely understood. Previous work has linked these correlations to
the bursty recurrence of topically relevant words. Here we generalise
that construction from individual lexical items to distributed semantic
observables. We map a text token by token onto a trajectory of
pre-trained word embeddings, use principal component analysis (PCA) to
identify text-specific directions of semantic variation, and threshold
the projected series to define inter-event times; their coefficients of
variation form a component-resolved burstiness spectrum $\{B_k\}$.
Across 28 literary and scientific texts and the first 20 components,
the originals remain systematically above the geometric
inter-event-time baseline and above two complementary controls: a
word-shuffled null, which preserves the embedding covariance, the PCA
eigensystem and the marginal projected distributions while destroying
temporal order, and a fractional-Gaussian-noise surrogate, which
preserves the Zipfian frequency structure and matches the detrended fluctuation analysis (DFA) exponent
of the original rank sequence. With events defined by the upper $5\%$
tail of each projection, the original exceeds both controls in every
text--component pair, while the two controls are
statistically indistinguishable from each other. Autocorrelation and
detrended fluctuation analysis of the semantic projections separate the
originals from both controls in the same way. Global rank-level
long-range correlations are therefore insufficient to generate the
bursty recurrence observed along semantic directions. The results
instead support an interpretation in terms of the ordered recurrence of
semantically related material.
\end{abstract}

\begin{keyword}
long-range correlations \sep burstiness \sep word embeddings \sep
principal component analysis \sep surrogate data
\end{keyword}

\end{frontmatter}

\section{Introduction}
\label{sec:intro}

\subsection{Long-range correlations in texts}

Written texts exhibit statistical structure over scales far beyond
local syntactic dependencies.
Among the most robust observations is the presence of long-range
correlations in symbolic sequences derived from natural language.
When a text is encoded as a numerical time series and analysed with
detrended fluctuation analysis (DFA) \citep{Peng1994}, the fluctuation
function commonly scales as $F(L)\sim L^\alpha$, with
$\alpha\in[0.6,0.8]$, above the uncorrelated value $\alpha=0.5$,
across languages, genres, and encoding choices
\citep{Schenkel1993,Ebeling1995,MontemurroPury2002}.

This observation raises a fundamental question: what aspect of textual
organisation produces memory over hundreds or thousands of tokens?
At least two statistical structures must be distinguished.
The first is static lexical composition, represented by the vocabulary
and word-frequency distribution.
Its most familiar regularity is Zipf's law \citep{Zipf1949}, according
to which the frequency $f(r)$ of the word of rank $r$ follows
$f(r)\sim r^{-\gamma}$ with $\gamma\simeq1$.
The second is the temporal ordering of words, through which syntax,
discourse, argument, and thematic development unfold.
A word-shuffled text preserves lexical composition and Zipf's law but
destroys temporal organisation, causing the DFA exponent to approach
$\alpha=0.5$ \citep{Ebeling1995,MontemurroPury2002}.

The global DFA exponent, however, does not by itself identify the
semantic mechanism responsible for long-range correlations.
The surrogate introduced in \citet{MontemurroDegliEsposti2026} maps fractional
Gaussian noise (FGN) onto the empirical rank-frequency structure and is
tuned to reproduce the DFA exponent of the original rank sequence.
It therefore provides a stationary, rank-level route to the same global
scaling and allows us to ask whether matching long-range memory in
frequency-rank space is sufficient to reproduce temporal organisation in
semantic space.

A geometric approach to long-range textual structure was introduced by
\citet{AlvarezLacalle2006}.
They constructed a co-occurrence representation within a sliding
``window of attention'' and used singular value decomposition to identify
text-specific concept vectors.
The motion of the sliding window through concept space produced
trajectories with power-law autocorrelation over long textual scales, and
surrogate tests connected this behaviour to hierarchical textual
organisation.
Their work established that semantic dynamics can be studied as motion
through a reduced concept space. More recently, \citet{GromovDang2023}
represented literary works in several languages as semantic trajectories
constructed from embeddings and $n$-grams, and analysed their dynamical
properties. These studies motivate treating a text as motion through
semantic space, while leaving open how recurrence and burstiness are
distributed across semantic directions.

Our approach shares this geometric viewpoint but differs in three
respects.
First, we use pre-trained static word embeddings rather than a
co-occurrence matrix estimated from a single text.
Second, the trajectory is resolved token by token rather than through an
aggregated sliding window.
Third, our primary observable is the burstiness of threshold crossings
along semantic directions, measured through the inter-event-time
coefficient of variation, together with the autocorrelation and DFA of
the projected series.

\subsection{From word-level burstiness to topical organisation}

A mechanism connecting topical organisation to long-range correlations
was developed by \citet{Altmann2012}.
Their framework treats language as a hierarchy of observables, from
topics to words and then to lower-level symbolic representations.
Correlations can propagate down this hierarchy, whereas burstiness is
visible most clearly at semantically relevant levels and can be lost
under projection to lower levels. Related statistical approaches have
shown that informative words are distributed non-uniformly and often
cluster strongly along a text
\citep{ChurchGale1995,Ortuno2002,Carpena2009,AltmannPierrehumbertMotter2009,TanakaIshiiBunde2016}, while
the scale-dependent inhomogeneity of word use carries measurable
semantic information \citep{MontemurroZanette2010}. Lexical cohesion
provides a complementary route to detecting subtopic structure from
changes in vocabulary across a document \citep{Hearst1997}.

Formally, a local matching condition $\mathcal{C}$ maps a text to a
binary sequence $u_t^{(\mathcal{C})}\in\{0,1\}$.
For example, $u_t^{(\mathcal{C})}=1$ may indicate that token $t$ is the
word ``prince''.
The gaps $\tau$ between successive events, i.e.\ between consecutive
positions with $u_t^{(\mathcal{C})}=1$, define an inter-event-time
(IET) distribution, whose coefficient of variation
\begin{equation}
  \CV=\frac{\sigmatau}{\meantau}
\end{equation}
measures temporal clustering.
Topically relevant words tend to occur in concentrated textual domains
and therefore show broad IET distributions and elevated burstiness,
whereas common function words are distributed more regularly even when
long-range correlations remain visible in lower-level observables.
This framework suggests a natural extension.
A topic such as war, philosophical reflection, geographical
distribution, or social interaction is not expressed by one keyword
alone.
It activates a distributed set of semantically related words.
A suitable topic-level observable should therefore respond to a region
or direction of semantic space rather than to the exact recurrence of a
single word type.

\subsection{Motivation and contribution}
 
We represent a text as an embedding trajectory
$(\mathbf{e}_t)_{t=1}^M$ in $\R^d$, where each retained token is mapped
to a pre-trained static word vector
\citep{Mikolov2013,Pennington2014}.
PCA of this trajectory identifies an ordered set of orthogonal,
text-specific directions of semantic variation.
Projecting the trajectory onto a direction $\mathbf{v}_k$ yields a
continuous semantic time series; thresholding that series produces a
binary event sequence to which the IET construction of
\citet{Altmann2012} can be applied.
The resulting spectrum $\{B_k\}_{k=1}^K$ measures burstiness separately
along the first $K$ principal semantic directions.
Throughout the paper we retain $K=20$; 
robustness to this choice is examined in Section~\ref{sec:single_text_robustness}.
 
This construction supports a component-resolved question:
how is semantic burstiness distributed across the principal directions
of a text?
The principal components are ordered by variance, but PCA does not
maximise burstiness.
The empirical relation between the eigenvalue spectrum
$\{\lambda_k\}$ and the burstiness spectrum $\{B_k\}$ therefore reveals
whether directions of strong semantic variation are also directions of
strong temporal clustering.
The analyses below show interpretable semantic contrasts along the
leading components, broad IET distributions in the original texts, and
elevated burstiness throughout the retained components.
They also show that the strength of association between semantic
variance and burstiness is text dependent: the two quantities are
related in some texts but are not interchangeable.
 
We then use two complementary null models to determine which statistical
properties produce this behaviour.
The word-shuffled null permutes the retained tokens, preserving the
static embedding geometry exactly while destroying temporal order.
The FGN surrogate retains the empirical rank-frequency structure and is
tuned to match the original rank-sequence DFA exponent, but generates
that scaling through a stationary rank-level process.
The comparison separates static semantic geometry from ordered semantic
dynamics, and global rank-level memory from burstiness in embedding
space.
 
Across the analysed corpus, both controls remain close to the geometric
IET baseline, whereas the original text is more bursty across essentially
all text--component pairs and over a broad range of threshold quantiles.
The projected autocorrelation functions and DFA exponents reinforce this
separation: semantic persistence remains visible in the original
trajectory but disappears from both controls, including the FGN
surrogate whose rank sequence retains the original global DFA exponent.
Thus matching lexical frequencies and rank-level long-range scaling is
not sufficient to reproduce semantic burstiness.

\subsection{Structure of the paper}

Section~\ref{sec:theory} develops the theoretical framework, including
semantic projections, PCA, the component-resolved burstiness spectrum,
and the two null models. Section~\ref{sec:pipeline} describes the
computational pipeline. Section~\ref{sec:oliver} presents the Results
for four representative texts, organised around semantic
interpretability, component-resolved burstiness, and threshold
robustness. Section~\ref{sec:corpus} establishes the corpus-level
generality of the component spectra and null-model separation.
Section~\ref{sec:temporal_persistence} examines persistence in semantic
projection space using autocorrelation and DFA. Section~\ref{sec:discussion} discusses implications
and limitations.

\section{Theoretical Framework}
\label{sec:theory}

\subsection{From binary indicators to continuous semantic projections}
\label{sec:observables}

Following \citet{Altmann2012}, let a text be a symbolic sequence
$\mathbf{s}=(s_1,s_2,\ldots,s_N)$ over a finite vocabulary
$\mathcal{A}$.
A local matching condition $\mathcal{C}$ defines a binary observable
\begin{equation}
  u_t^{(\mathcal{C})}=\mathbf{1}[s_t\text{ satisfies }\mathcal{C}].
  \label{eq:binary_indicator}
\end{equation}
The autocorrelation of this sequence and the diffusion exponent of the
cumulative count $X(t)=\sum_{j=1}^t u_j^{(\mathcal{C})}$, defined by
$\sigma_X^2(t)\sim t^{2H}$, characterise its long-range structure.
Here $H$ coincides with the DFA exponent $\alpha$ of the stationary
series $u_t^{(\mathcal{C})}$, the quantity used throughout this paper.
We generalise the local lexical condition to a distributed semantic
observable.
Let $\phi:\mathcal{A}\to\R^d$ be a pre-trained static word embedding.
After preprocessing and embedding lookup, the retained tokens are
relabeled consecutively as $(\tilde s_1,\ldots,\tilde s_M)$, with
$M\leq N$.
Their \emph{embedding trajectory} is
\begin{equation}
  \mathcal{E}=(\et)_{t=1}^M,
  \qquad
  \et=\phi(\tilde s_t).
  \label{eq:trajectory}
\end{equation}

For a unit vector $\mathbf{v}\in\R^d$, the corresponding
\emph{semantic projection} is
\begin{equation}
  x_t^{\mathbf{v}}=(\et-\ebar)\cdot\mathbf{v},
  \qquad
  \ebar=\frac{1}{M}\sum_{t=1}^M\et.
  \label{eq:projection}
\end{equation}
At threshold $\theta$, this continuous series defines the binary event
sequence
\begin{equation}
  b_t^{\mathbf{v}}(\theta)
  =\mathbf{1}\!\left[x_t^{\mathbf{v}}>\theta\right].
  \label{eq:binary_from_proj}
\end{equation}
The projection describes motion along a semantic direction, while the
binary series records visits to its upper tail and supplies the events
whose recurrence statistics are analysed.

\begin{remark}
  If $\mathbf{v}$ is parallel to $\phi(w)-\ebar$ for a word type $w$,
  then $x_t^{\mathbf{v}}$ tends to be large when the current token is
  $w$ and for tokens whose embeddings are close to, or strongly aligned
  with, $\phi(w)$.
  Thresholding this projection therefore gives a softened semantic
  analogue of the exact word indicator in
  Eq.~\eqref{eq:binary_indicator}.
  The construction extends the unit of analysis from one lexical item to
  a distributed semantic neighbourhood.
\end{remark}

\subsection{PCA of the embedding trajectory}
\label{sec:pca}

The direction $\mathbf{v}$ determines the semantic observable.
We use the principal components of the embedding trajectory to obtain a
text-specific orthonormal basis ordered by captured variance.
The empirical covariance matrix of the mean-centred trajectory is
\begin{equation}
  \Sig=
  \frac{1}{M}\sum_{t=1}^M
  (\et-\ebar)(\et-\ebar)^\top
  \in\R^{d\times d}.
  \label{eq:covariance}
\end{equation}
Its eigendecomposition
\begin{equation}
  \Sig\vk=\lk\vk,
  \qquad
  \lambda_1\geq\lambda_2\geq\cdots\geq\lambda_d\geq0,
  \label{eq:pca_eigenproblem}
\end{equation}
yields the principal directions $\vk$ and their variances $\lk$.
The projected series for component $k$ is
\begin{equation}
  \xtk=(\et-\ebar)\cdot\vk,
  \qquad t=1,\ldots,M.
  \label{eq:proj_series}
\end{equation}

PCA directions maximise variance, not semantic interpretability or
burstiness.
Their semantic meaning is assessed empirically from the word types at
the positive and negative poles of each component.
The ordered eigenvalue spectrum describes the static distribution of
semantic variance, while the temporal statistics of $\xtk$ describe how
the text moves along these directions.

A key property of $\Sig$ is its invariance under permutation of the
token sequence.
For a permutation $\pi$ of $\{1,\ldots,M\}$ the mean is unchanged,
$\ebar^{\pi}=\ebar$, and therefore
\begin{equation}
  \Sig^{\pi}
  =\frac{1}{M}\sum_{t=1}^M
  \bigl(\mathbf{e}_{\pi(t)}-\ebar\bigr)
  \bigl(\mathbf{e}_{\pi(t)}-\ebar\bigr)^\top
  =\Sig.
  \label{eq:sigma_invariant}
\end{equation}

It follows that a word-shuffled surrogate has the same eigenvalues,
eigenvectors, and marginal distribution of every projected series as
the original. PCA therefore defines a static geometry of the embedded lexical
composition, whereas burstiness, autocorrelation, and DFA probe the
ordering of visits through that geometry.

\subsection{The burstiness spectrum}
\label{sec:burstiness_spectrum}

For component $k$, the projected series is thresholded at its empirical
$q$-quantile:
\begin{equation}
  b_t^{(k)}=
  \mathbf{1}\!\left[
  \xtk>Q_q\!\left(\{\xtk\}_{t=1}^M\right)
  \right].
  \label{eq:binary_k}
\end{equation}
The event rate is $p=1-q$ up to discretisation.
Let $\tau_1^{(k)},\tau_2^{(k)},\ldots$ be the gaps between successive
positions for which $b_t^{(k)}=1$.
The burstiness of component $k$ is measured by the coefficient of
variation of the IET sequence, following \citet{Altmann2012}:
\begin{equation}
  \bk
  =\CV\!\left(\tau^{(k)}\right)
  =\frac{\sigmatau^{(k)}}{\meantau^{(k)}}.
  \label{eq:burstiness_def}
\end{equation}
The collection $\{\bk\}_{k=1}^K$ is the \emph{burstiness spectrum} of
the text.
We use the unnormalised coefficient of variation rather than the
bounded burstiness parameter of \citet{GohBarabasi2008},
$B^{\rm GB}=(\CV-1)/(\CV+1)\in[-1,1]$, since the former compares
directly with the finite-rate reference derived below; the two are
related by a monotone transformation and induce the same ordering.

At finite event rate, the appropriate independent-event reference is a
geometric distribution rather than an exponential distribution.
For an i.i.d. Bernoulli$(p)$ event sequence,
\begin{equation}
  P(\tau=n)=(1-p)^{n-1}p,
  \qquad n\geq1,
\end{equation}
which gives
\begin{equation}
   \CV_{\rm geom}(p)=\sqrt{1-p}=\sqrt{q}.
  \label{eq:cv_geometric}
\end{equation}
For the default $q=0.95$, $p=0.05$ and
$\CV_{\rm geom}=\sqrt{0.95}\simeq0.975$.
\citet{Altmann2012} adopt the continuous-time convention $\CV>1$; at
the event rates used here the discrete correction matters, since
independent events at $q=0.95$ give $0.975$ rather than $1$.
Values of $B_k$ above this reference indicate more irregular,
clustered recurrence than expected from independent events at the same
rate. A derivation of the discrete geometric reference is given in
Section~\ref{app:geometric} of the Supplementary Material.

Because all three sequences are permutations of the same embedded token
multiset, their projected series share an identical marginal
distribution. The empirical $q$-quantile and resulting event count are
therefore identical across the three sequences; the event rate is
$1-q$ up to finite-sample and tie discretisation.
\subsection{Semantic variance and burstiness across components}
\label{sec:component_structure}

The PCA index $k$ orders directions by decreasing variance, 
whereas the same index carries no ordering of $B_k$, which need not be monotone in $k$.
The pair of spectra
\begin{equation}
  \{\lambda_k\}_{k=1}^K,
  \qquad
  \{B_k\}_{k=1}^K
  \label{eq:two_spectra}
\end{equation}
therefore separates two properties of the text: the amount of semantic
variation along a direction and the temporal clustering of visits along
that direction.
The component-resolved analysis asks whether burstiness is elevated
above the geometric and surrogate baselines, how it varies across the
retained PCA directions, and how strongly it is associated with
semantic variance.

We quantify the last relation by the Pearson correlation
\begin{equation}
  r(\lambda,B)=
  \frac{\sum_{k=1}^{K}(\lambda_k-\bar\lambda)(B_k-\bar B)}
  {\sqrt{\sum_{k=1}^{K}(\lambda_k-\bar\lambda)^2}
   \sqrt{\sum_{k=1}^{K}(B_k-\bar B)^2}},
  \label{eq:eigen_burst_corr}
\end{equation}
where $\bar\lambda$ and $\bar B$ denote component means over the retained
range.

The $K$ pairs $(\lambda_k,B_k)$ are not independent draws: $\{\lambda_k\}$
is ordered by construction and both spectra derive from the same
trajectory. We therefore do not use the parametric significance test for
$r$. Significance is assessed by a permutation test over the pairing
between the two spectra ($10^5$ random pairings), and we report the
Spearman rank correlation $\rho$ alongside $r$, since the eigenvalue
spectrum is strongly skewed and $r$ is otherwise dominated by the leading
components.

Throughout, $P$ denotes a significance level, to avoid confusion with
the event rate $p=1-q$.

A positive value indicates that high-variance directions tend also to
be more bursty, but even a positive correlation does not make the two spectra interchangeable:
a lower-variance component can be highly bursty if it is
associated with a topic concentrated in a limited number of textual
regions.
We report both the full burstiness spectrum and its relation to the eigenvalue spectrum.

\subsection{The word-shuffled null model}
\label{sec:null_model_shuffle}

The word-shuffled null tests whether elevated burstiness can be
explained by static lexical composition and embedding geometry alone.
Given the retained token sequence $(\tilde s_1,\ldots,\tilde s_M)$, a
surrogate is formed by a uniformly random permutation
$\pi\in S_M$.
Its embedding trajectory is therefore a permutation of the original
trajectory. Because a single permutation is itself a random draw, the reported
word-shuffled quantities are averaged over $N_{\rm surr}$ independent
permutations, the spread across realisations giving the mean$\pm2\sigma$
bands shown in the figures. The burstiness spectrum and the DFA
exponents are averaged in this way; the autocorrelation is reported for
a single realisation, which suffices because the shuffled autocorrelation function (ACF) is
indistinguishable from zero at every lag.

By Eq.~\eqref{eq:sigma_invariant}, the surrogate preserves exactly the
word counts of the retained sequence, the embedding covariance matrix
and complete PCA eigensystem, and the marginal distribution of every
projected series $\xtk$.
It destroys temporal order.
Consequently, differences in IET distributions, burstiness,
autocorrelation, or DFA arise from the ordering of visits through
semantic space rather than from vocabulary, frequency distribution, or
static embedding geometry.

\subsection{The FGN surrogate as a long-range null model}
\label{sec:null_model_fgn}

The FGN surrogate of \citet{MontemurroDegliEsposti2026} provides a complementary
null model.
Starting from the Zipf-rank sequence
$\mathbf{r}=(r_1,\ldots,r_N)$, a fractional Gaussian noise process is
mapped onto the empirical rank-frequency structure.

Its generator exponent $\alpha_0$ is chosen by bisection so that the DFA
exponent measured on the surrogate rank sequence matches that of the
original,
\begin{equation}
  \alpha(\mathbf{r}^{\rm FGN})\simeq\alpha(\mathbf{r}),
  \label{eq:fgn_dfa_match}
\end{equation}
where $\alpha_0>\alpha$ in general because the rank-to-word mapping is
not exponent-preserving; Table~\ref{tab:corpus} reports both.

The resulting ranks are mapped back to words and processed through the
same filtering, embedding, and projection pipeline as the original.
Because the rank-to-word mapping reproduces the empirical rank histogram
exactly, the resulting word sequence is a permutation of the original
token sequence. Applying the same deterministic frequency filter and
embedding lookup therefore gives the same retained token multiset and
the same trajectory length $M$.

The construction preserves the raw token count $N$ and empirical
frequency structure and reproduces the rank-level DFA exponent by
design. Although it is also an exact permutation, unlike the
word-shuffled null its ordering is generated to reproduce long-range
rank-level scaling rather than by uniform randomisation. It therefore
tests whether stationary correlations between word ranks are sufficient
to produce the burstiness observed along semantic projections.

The exponent is matched on the unfiltered rank sequence of length $N$,
whereas burstiness, autocorrelation and DFA of the projections are
measured after frequency filtering on a trajectory of length $M$.
The same filter, derived from the original frequency table, is applied
to the original and both controls.

Because the frequency filter removes the most frequent word types, which
occupy a large share of the positions of the sequence, it need not
affect the rank-level memory of the original and of the FGN surrogate in
the same way. Because DFA estimates depend on the range of scales used
in the fit \citep{Kantelhardt2001}, we measure this with the estimator
of the matching step. Applied to the $M$ retained tokens and with the ranks of the original frequency table,
the rank-sequence DFA exponent falls on average from $0.733$ to $0.670$
for the original texts and to $0.629$ for the FGN surrogates: the
filtered surrogate retains a weaker memory than the filtered original
in 22 of the 28 texts (mean difference $-0.040$, paired $t=-3.7$,
$P<10^{-3}$), although it remains well above the value $0.49$ of the
shuffled sequence (SI, Section~\ref{app:filtered_fgn}). To exclude that
the control is weakened by this loss, we also construct an FGN surrogate
matched after filtering, FGN$_M$: it is generated directly on the $M$
retained tokens, with $\alpha_0$ tuned so that the DFA exponent of its
rank sequence equals that of the original retained tokens. FGN$_M$ is
again a permutation of the same multiset; its results are reported in
Section~\ref{sec:corpus_diffs}.

\begin{remark}
\label{rem:fgn_cov}
  Since the three sequences are permutations of the same token multiset,
  Eq.~\eqref{eq:sigma_invariant} implies that they have exactly the same
  embedding covariance matrix, complete PCA eigensystem, and marginal
  distribution of every projected series. Projecting both controls onto
  the original PCA directions is therefore exact rather than an
  approximation.
\end{remark}

\begin{remark}
\label{rem:rank_semantic}
  Rank proximity is not semantic proximity.
  Words of similar frequency can occupy widely separated regions of
  embedding space, so correlations imposed on the rank sequence need
  not generate correlations between the corresponding embedding
  vectors.
  The FGN surrogate tests this distinction directly: it retains global
  rank-level scaling without explicitly organising semantically related
  words into the same textual regions.
\end{remark}

\subsection{Logical role of the null models}
\label{sec:taxonomy}

Both controls follow the logic of surrogate-data testing
\citep{Theiler1992,SchreiberSchmitz2000}: each preserves a specified set
of properties of the original sequence and randomises everything else,
so that a statistic separating the original from the surrogate excludes
the corresponding null hypothesis.
The three sequences differ in which statistical ingredients they retain;
Table~\ref{tab:null_taxonomy} summarises the differences.

\begin{table}[tbp]
\centering
\caption{Statistical ingredients retained by each sequence.}

\label{tab:null_taxonomy}
\small
\begin{tabular}{lccc}
\toprule
Ingredient & Original & Word-shuffled & FGN \\
\midrule
Lexical composition, frequency structure & present & exact & exact \\
Static embedding geometry ($\Sig$, eigensystem) & present & exact & exact \\
Temporal ordering in embedding space & present & destroyed & destroyed \\
Rank-level long-range scaling & present & destroyed & matched \\
\bottomrule
\end{tabular}
\end{table}

The comparisons therefore address two distinct questions.
First, does the ordered original trajectory produce component-level
burstiness that is absent when the same embedded tokens are randomly
permuted?
Second, is matching the global DFA exponent of the rank sequence
sufficient to reproduce that semantic burstiness?
The Results answer both by comparing IET distributions, burstiness
spectra, autocorrelation functions and projection-level DFA exponents
across the three sequences.

Both answers are negative, and the two negatives are not redundant:
neither the static geometry of the embedded vocabulary nor global
rank-level memory is sufficient to generate the observed
component-level burstiness. The comparison therefore supports the
interpretation that the ordering of visits through semantic space is an
essential ingredient of the observed effect.

\section{Computational Pipeline}
\label{sec:pipeline}

The following sections describe the implementation of the framework
introduced in Section~\ref{sec:theory}.

\subsection{Text preprocessing}
\label{sec:preprocessing}
\label{sec:filtering_results}

Input texts are the plain-text editions of Project Gutenberg. For each
work, the body of the text is extracted between two word anchors
recorded in a manifest distributed with the code: it runs from the
first chapter, or the author's own introduction, to the end of the last
chapter, epilogue or conclusion, so that editorial prefaces, tables of
contents, lists of illustrations, appendices, indices and endnotes are
excluded. The body is then normalised: typographic quotation marks and
apostrophes are mapped to ASCII, diacritics are removed (so that, for
instance, transliterated names printed with stress accents are found in
the embedding vocabulary), illustration, footnote and image markup and
chapter labels are removed, the abbreviations \emph{Mr.}, \emph{Mrs.},
\emph{Dr.} and \emph{St.} are kept as words, and hyphenated compounds
are split into their components.
Tokenisation uses the NLTK \texttt{word\_tokenize} function (version
3.9.1) applied to the lowercased text, retaining only alphabetic tokens;
punctuation, numerals and clitic fragments such as \emph{n't} are
thereby discarded.

Two frequency-based filters are then applied.
Let $\mathrm{rank}(w)$ denote the rank of word $w$ in the vocabulary
ordered by decreasing frequency.
\begin{enumerate}
  \item \emph{High-frequency removal}: words with
        $\mathrm{rank}(w) \leq R$ are removed.
        Default: $R = 100$.
  \item \emph{Hapax removal}: words with frequency $f(w) < F_{\min}$
        are removed.
        Default: $F_{\min} = 5$.
\end{enumerate}
The motivation for the high-frequency filter deserves emphasis.
Without it, the dominant PCA direction $\mathbf{v}_1$ of the embedding
trajectory is strongly dominated by a frequency contrast: the negative
pole concentrates high-frequency function words (\emph{the, of, and,
to, a, ...}), while the positive pole concentrates rarer content words
whose embedding vectors are geometrically distant from the
high-frequency cluster. That the leading principal directions of
pre-trained word embeddings encode word frequency is a known property
of these representations \citep{MuViswanath2018}. Removing the top-$R$ most frequent words
suppresses this dominant function-word contrast and makes topical
structure more visible in the leading components, but it does not imply
that PC1 itself is a purely topical direction.

The choice $R=100$ is empirically motivated by direct inspection of word
scores along the leading components and is used consistently throughout
the reported analyses.

A known residual artefact persists in PC1 for texts whose vocabulary
contains many rare transliterated proper nouns or archaic forms
(e.g.\ \emph{War and Peace}): these words
cluster in a peripheral region of the GloVe embedding space and
partially contaminate PC1 regardless of $R$.
This artefact is independent of topical structure and is documented
where relevant. Component spectra retain PC1 for completeness, but its
semantic interpretation and short-lag temporal statistics are treated
cautiously; results for $k \geq 2$ are unaffected by this particular
artefact.

\subsection{Word embedding model}
\label{sec:embedding}

We use GloVe vectors trained on Wikipedia and Gigaword
\citep{Pennington2014}, with $d = 300$ dimensions and a vocabulary of
approximately $400{,}000$ words, in the \texttt{glove.6B.300d}
release of Stanford NLP; its SHA-256 checksum is verified when the
vectors are loaded. The \texttt{glove-wiki-\allowbreak gigaword-300}
model distributed through \texttt{gensim.\allowbreak downloader}
contains the same vectors.

The choice of static (non-contextual) embeddings is deliberate.
For static embeddings, $\et = \phi(\tilde s_t)$ depends only on the word
\emph{type} $s_t$, making the trajectory a deterministic function of
the token sequence.
Shuffling the token sequence thus has an unambiguous effect: it
permutes the trajectory while leaving the multiset $\{\et\}$ unchanged
--- as does the FGN surrogate (Remark~\ref{rem:fgn_cov}).
Contextual embeddings (BERT, GPT) would assign different vectors to
the same word type depending on local context, introducing dependencies
between the embedding and the ordering that would complicate the
surrogate comparison.
We defer contextual embedding analysis to future work
(Section~\ref{sec:discussion}).

Out-of-vocabulary tokens are skipped; their original positions are not
used in the trajectory or in IET computations.
The OOV rate after filtering is below $2\%$ for the four representative
texts (Table~\ref{tab:single_text_summary}), consisting mainly of
proper nouns, archaic forms, and dialect words.

\subsection{PCA and projection}
\label{sec:pca_pipeline}

The embedding trajectory is assembled as a matrix
$\mathbf{V} \in \R^{M \times d}$ with rows $\et^\top$.
After mean-centring, $\mathbf{X} = \mathbf{V} - \mathbf{1}\ebar^\top$,
PCA is computed from the sample covariance matrix by symmetric
eigendecomposition in double precision. Each eigenvector is oriented so
that its entry of largest absolute value is positive (the convention of
\texttt{scikit-learn}); the orientation matters because events are
defined by the upper tail of the projections.
The full eigenvalue spectrum $\{\lambda_k\}_{k=1}^{d}$ is retained for
the spectral analysis, while eigenvectors and projections are formed for
the first $K = 20$ components, giving the projected series matrix
$\mathbf{P} \in \R^{M \times K}$, $P_{tk} = \xtk$, used in all
subsequent analyses.

\subsection{FGN surrogate generation}
\label{sec:fgn_pipeline}

The FGN surrogate is constructed using the pipeline of
\citet{MontemurroDegliEsposti2026}:
\begin{enumerate}
  \item The unfiltered token stream is tokenised and each token is
        replaced by its Zipf rank, producing a rank sequence
        $\mathbf{r}$ of length $N$.
  \item The DFA exponent $\alpha$ of the original text is estimated
        from $\mathbf{r}$.
  \item A FGN process with parameter $\alpha_0$ is generated and
        mapped onto the empirical rank histogram.
        The parameter $\alpha_0$ is found by bisection so that the
        DFA exponent of the surrogate matches $\alpha$.
  \item The surrogate rank sequence is mapped back to words via the
        inverse Zipf mapping. This reproduces the empirical rank
        histogram, so the resulting word sequence
        $\mathbf{s}^{\rm FGN}$ is a permutation of the original.
\end{enumerate}
The resulting word sequence is then processed through the same
frequency filter (Section~\ref{sec:preprocessing}), embedding lookup,
and projection pipeline as the original text, and projected onto the
original PCA eigenvectors $\{\vk\}$. Because $\mathbf{s}^{\rm FGN}$ is a
permutation of the original token sequence, the filtered surrogate has
the same retained token multiset and length $M$, and
$\Sig_{\rm FGN}=\Sig_{\rm orig}$ exactly
(Remark~\ref{rem:fgn_cov}): the original eigenvectors \emph{are} the
surrogate's own principal directions, and projecting onto them involves
no approximation.

\section{Results: semantic directions and burstiness spectra}
\label{sec:oliver}

\subsection{Representative texts and summary}
\label{sec:representative_summary}

We first examine four texts chosen to span distinct forms of semantic
and narrative organisation: a long polyphonic novel (\emph{War and
Peace}), a scientific treatise (\emph{On the Origin of Species}), a
Victorian novel of formation (\emph{Great Expectations}), and a
modernist novel with marked episode-to-episode stylistic variation
(\emph{Ulysses}). All analyses use the pipeline of
Section~\ref{sec:pipeline}, with $R=100$, $F_{\min}=5$, $q=0.95$, and
$K=20$ components unless stated otherwise.

The Results are organised by empirical question rather than by book.
We first establish that the PCA basis defines interpretable,
text-specific directions while remaining a static property of lexical
composition. We then examine the inter-event-time distributions and the
component-resolved burstiness spectra, followed by robustness to the
threshold quantile. Table~\ref{tab:single_text_summary} collects the
main numerical descriptors.

\begin{table}[tbp]
\centering
\caption{Summary statistics for the four representative texts.
$N$: raw tokens; $M$: tokens after filtering; OOV: fraction of filtered
tokens absent from the embedding vocabulary; $\alpha$: DFA exponent of the
rank sequence of the full token stream; $\alpha_0$: generator exponent of the matched FGN
surrogate; $r$ and $\rho$: Pearson and Spearman correlation between
$\lambda_k$ and $B_k$ over $k = 1,\ldots,20$; $B_{\max}$: largest
burstiness value, attained at component $k_{\max}$.
Significance from the permutation test over component pairings:
$^{*}P<0.05$, $^{**}P<0.01$, $^{***}P<0.001$; no mark: $P>0.05$.}
\label{tab:single_text_summary}
\small
\begin{tabular}{@{}l cccc@{}}
\toprule
 & \emph{War and} & \emph{Origin of} & \emph{Great} & \\
 & \emph{Peace}   & \emph{Species}   & \emph{Expectations} & \emph{Ulysses} \\
\midrule
Raw tokens $N$                & 564{,}609 & 150{,}812 & 186{,}170 & 263{,}007 \\
Filtered tokens $M$           & 235{,}426 &  58{,}583 &  62{,}754 & 100{,}539 \\
OOV fraction                  & 1.6\%     & 0.4\%     & 1.3\%     & 0.7\%     \\
DFA exponent $\alpha$         & 0.724     & 0.772     & 0.703     & 0.778     \\
FGN generator $\alpha_0$      & 0.76      & 0.80      & 0.76      & 0.80      \\
Pearson $r(\lambda,B)$        & 0.708$^{**}$  & 0.393        & 0.421        & 0.664$^{*}$   \\
Spearman $\rho(\lambda,B)$    & 0.847$^{***}$ & 0.678$^{**}$ & 0.611$^{**}$ & 0.887$^{***}$ \\
Maximum burstiness $B_{\max}$ & 2.019     & 1.876     & 1.249     & 1.470     \\
Component $k_{\max}$          & 3         & 2         & 3         & 2         \\
\bottomrule
\end{tabular}

\end{table}

Across the four texts, the original projection sequences are more bursty
than both controls for all $K=20$ components at $q=0.95$. Semantic
variance and temporal clustering are monotonically associated in every
text ($\rho$ from $0.611$ to $0.887$, significant in all four) but are
not interchangeable: the linear correlation is weaker ($r$ from $0.393$
to $0.708$) and significant only for \emph{War and Peace} and
\emph{Ulysses}, and the burstiest direction is never the first
principal component, falling at PC3, PC2, PC3 and PC2 respectively.
These component indices are descriptive rather than invariant
identifiers: when neighbouring eigenvalues are close, small rotations
of the corresponding PCA axes can change an individual $B_k$. The
stability test in Supplementary Section~\ref{app:bk_stability} shows
that the corpus-level quantities are much less sensitive to this
effect.

\subsection{Static semantic geometry and interpretable directions}
\label{sec:semantic_geometry_results}

The eigenvalue spectra describe the geometry of the retained embedded
vocabulary. As established in Section~\ref{sec:pca}, word shuffling
leaves this geometry exactly invariant, and so does the FGN surrogate,
which is also a permutation of the original token sequence. The semantic poles of the
leading components provide a separate empirical test of whether those
directions correspond to interpretable contrasts.

Taken together, the four texts show that PCA produces an ordered static
geometry whose leading directions are only partly text-specific. Each
low-order component pairs a pole that is characteristic of the
individual work with a pole that recurs across texts: the positive pole
of PC2 is military in \emph{War and Peace}, riverine in \emph{Great
Expectations}, geographical in \emph{On the Origin of Species} and
titular in \emph{Ulysses}, whereas its negative pole is the same
colloquial register (\emph{really}, \emph{yourself}, \emph{feel}) in
every novel. PC1 is common to all four texts in both poles and reflects
the geometry of the embedding space rather than the organisation of any
individual work. The temporal question is separate: whether visits
along these directions occur regularly or in concentrated textual
episodes.

\subsubsection{\emph{War and Peace}}
\label{sec:wrnpc}

\emph{War and Peace} is the longest representative text
($N=564{,}609$ raw tokens). After filtering and embedding lookup, the
trajectory contains $M=235{,}426$ tokens; the rank-sequence DFA exponent
is $\alpha=0.724$, and the fitted FGN parameter is $\alpha_0=0.76$.
The corresponding eigenvalue spectrum, shown in
Figure~\ref{fig:wrnpc_spectra} of the Supplementary Material, gives
$7.00\%$ and $4.21\%$ of the variance for PC1 and PC2, respectively,
with the first ten components accounting for $27.2\%$.

The semantic poles in Table~\ref{tab:wrnpc_pc} reveal two especially clear contrasts. PC2
separates military vocabulary (\emph{regiment}, \emph{infantry},
\emph{cavalry}, \emph{battalion}) from personal and emotional discourse
(\emph{really}, \emph{yourself}, \emph{feel}, \emph{understand}). PC3
contrasts physical action and embodied presence with philosophical and
moral reflection. PC1 partly reflects the known GloVe artefact in which
transliterated Russian names and emotionally inflected rare words lie
at the periphery of the embedding cloud; it is retained in the spectra
but interpreted cautiously.

\begin{table}[tbp]
\centering
\caption{Top-6 word types at the positive and negative poles of PC1,
PC2 and PC3 for \emph{War and Peace} ($R=100$, $F_{\min}=5$),
with frequencies in the filtered token sequence.
$B_k$ is a property of the component, not of the pole, and is therefore
shown once per pair; the largest value over the $K=20$ components is
shown in bold.}
\label{tab:wrnpc_pc}
\small
\setlength{\tabcolsep}{4pt}
\begin{tabular}{llr p{5.4cm} p{4.4cm}}
\toprule
PC & Pole & $B_k$ & Top words (freq) & Interpretation \\
\midrule
PC1 & $+$ & 1.637 &
  \emph{gerasim} (26), \emph{maidservant} (5), \emph{irritably} (6),
  \emph{piteously} (7), \emph{nikolaevich} (6), \emph{andreevich} (8)
  & Rare transliterated names and low-frequency emotive vocabulary
    (GloVe artefact) \\
PC1 & $-$ & &
  \emph{any} (466), \emph{government} (64), \emph{than} (533),
  \emph{its} (568), \emph{people} (557), \emph{us} (423)
  & High-frequency function words and abstract nouns (GloVe artefact) \\
\midrule
PC2 & $+$ & 1.875 &
  \emph{regiment} (227), \emph{infantry} (82), \emph{cavalry} (77),
  \emph{battalion} (44), \emph{squadron} (64), \emph{brigade} (5)
  & Military units and formations (War) \\
PC2 & $-$ & &
  \emph{really} (203), \emph{yourself} (117), \emph{feel} (133),
  \emph{myself} (148), \emph{laugh} (60), \emph{everybody} (119)
  & Personal and emotional discourse (Peace) \\
\midrule
PC3 & $+$ & \textbf{2.019} &
  \emph{walked} (70), \emph{sat} (345), \emph{feet} (112),
  \emph{screaming} (6), \emph{yards} (20), \emph{clutching} (12)
  & Physical action and embodied presence \\
PC3 & $-$ & &
  \emph{principles} (7), \emph{undertake} (10), \emph{utmost} (12),
  \emph{renounce} (8), \emph{justification} (11), \emph{obligations} (7)
  & Moral obligation and abstract reasoning \\
\bottomrule
\end{tabular}
\end{table}

\subsubsection{\emph{On the Origin of Species}}
\label{sec:darwin}

\emph{On the Origin of Species} contains $N=150{,}812$ raw tokens and
$M=58{,}583$ retained tokens, with $\alpha=0.772$ and
$\alpha_0=0.80$. PC1 explains $6.64\%$ of the variance, PC2
$3.61\%$, and the first ten components $26.4\%$.

PC1 is dominated by specialised nineteenth-century biological and
geological terminology at one pole and pronouns or conversational
markers at the other, and therefore contains a register-related
embedding artefact. PC2 is a substantive geographical-versus-epistemic
direction: spatial and biogeographical terms are opposed to the
language of inference and theoretical reasoning. PC3 contrasts
first-person field observations and concrete specimens with abstract
classification and synthesis. The corresponding poles are listed in
Table~\ref{tab:darwin_pc}.

\begin{table}[tbp]
\centering
\caption{Top-6 word types at the positive and negative poles of PC1,
PC2 and PC3 for \emph{On the Origin of Species} ($R=100$,
$F_{\min}=5$), with frequencies in the filtered token sequence.
$B_k$ is a property of the component, not of the pole, and is therefore
shown once per pair; the largest value over the $K=20$ components is
shown in bold.}
\label{tab:darwin_pc}
\small
\setlength{\tabcolsep}{4pt}
\begin{tabular}{llr p{5.4cm} p{4.4cm}}
\toprule
PC & Pole & $B_k$ & Top words (freq) & Interpretation \\
\midrule
PC1 & $+$ & 1.250 &
  \emph{sessile} (11), \emph{faunas} (14), \emph{ovules} (6),
  \emph{plumage} (9), \emph{polymorphic} (9), \emph{insectivorous} (6)
  & Specialised 19th-c.\ natural history lexicon \\
PC1 & $-$ & &
  \emph{he} (163), \emph{him} (17), \emph{said} (47),
  \emph{us} (104), \emph{she} (24), \emph{his} (90)
  & Pronouns and interpersonal register (GloVe artefact) \\
\midrule
PC2 & $+$ & \textbf{1.876} &
  \emph{northern} (37), \emph{islands} (131), \emph{sea} (100),
  \emph{southern} (43), \emph{island} (60), \emph{eastern} (13)
  & Places and biogeographical setting \\
PC2 & $-$ & &
  \emph{innate} (7), \emph{slightest} (7), \emph{understand} (74),
  \emph{infer} (31), \emph{absolutely} (25), \emph{judgment} (12)
  & Scientific reasoning and theoretical inference \\
\midrule
PC3 & $+$ & 1.740 &
  \emph{me} (153), \emph{eat} (5), \emph{squirrels} (5),
  \emph{beaks} (10), \emph{legs} (37), \emph{rabbit} (5)
  & Field observations, specific specimens, first person \\
PC3 & $-$ & &
  \emph{geographical} (42), \emph{reduction} (5),
  \emph{increased} (27), \emph{proportional} (16),
  \emph{increase} (66), \emph{increases} (6)
  & Quantitative comparison and proportional reasoning \\
\bottomrule
\end{tabular}
\end{table}

\subsubsection{\emph{Great Expectations}}
\label{sec:greatexp}

\emph{Great Expectations} contains $N=186{,}170$ raw tokens and
$M=62{,}754$ retained tokens, with $\alpha=0.703$ and
$\alpha_0=0.76$. PC1, PC2, and the first ten components account for
$6.22\%$, $4.46\%$, and $27.1\%$ of the variance, respectively.

The positive pole of PC1 contains character names and Victorian
dialect, opposed to generic and abstract vocabulary, and is treated as
an embedding artefact. PC2 contrasts the spatial and kinetic language
of the river and the open country with introspective and conversational
language. PC3 opposes bodily action and gesture to a formal Latinate
register. The corresponding poles are listed in
Table~\ref{tab:greatexp_pc}.

\begin{table}[tbp]
\centering
\caption{Top-6 word types at the positive and negative poles
of PC1, PC2, and PC3 for \emph{Great Expectations}
($R=100$, $F_{\min}=5$).}
\label{tab:greatexp_pc}
\small
\setlength{\tabcolsep}{4pt}
\begin{tabular}{llr p{5.4cm} p{4.4cm}}
\toprule
PC & Pole & $B_k$ & Top words (freq) & Interpretation \\
\midrule
PC1 & $+$ & 1.189 &
  \emph{jaggers} (242), \emph{townsman} (6), \emph{biddy} (231),
  \emph{peeped} (5), \emph{stammered} (6), \emph{provis} (70)
  & Character names and Victorian dialect (GloVe artefact) \\
PC1 & $-$ & &
  \emph{will} (174), \emph{government} (5), \emph{its} (151),
  \emph{year} (26), \emph{states} (6), \emph{their} (175)
  & High-frequency generic and abstract vocabulary \\
\midrule
PC2 & $+$ & 1.217 &
  \emph{river} (59), \emph{near} (66), \emph{opened} (41),
  \emph{fell} (55), \emph{port} (7), \emph{centre} (5)
  & Spatial and kinetic language (river, marshes, forge) \\
PC2 & $-$ & &
  \emph{really} (41), \emph{understand} (53), \emph{think} (241),
  \emph{anybody} (30), \emph{myself} (236), \emph{yourself} (51)
  & Introspective and conversational register \\
\midrule
PC3 & $+$ & \textbf{1.249} &
  \emph{walked} (45), \emph{legs} (35), \emph{feet} (34),
  \emph{staring} (26), \emph{crying} (15), \emph{bed} (72)
  & Bodily action and gesture \\
PC3 & $-$ & &
  \emph{concerning} (13), \emph{regarding} (8), \emph{recognised} (14),
  \emph{expressly} (5), \emph{derived} (13), \emph{relation} (12)
  & Formal Latinate register \\
\bottomrule
\end{tabular}
\end{table}

\subsubsection{\emph{Ulysses}}
\label{sec:ulysses}

\emph{Ulysses} contains $N=263{,}007$ raw tokens and $M=100{,}539$
retained tokens, with $\alpha=0.778$ and $\alpha_0=0.80$. Its first
ten components account for $25.1\%$ of the variance, the smallest
fraction among the four texts, consistent with the diversity of
registers across its eighteen episodes.

PC1 is strongly affected by Dublin-specific proper names,
multilingual fragments, and neologisms. PC2 separates formal titles and
institutional naming from colloquial and intimate discourse. PC3
contrasts physical and material description with the vocabulary of
kinship, marriage and literary identity. These semantic poles are
listed in Table~\ref{tab:ulysses_pc}.

\begin{table}[tbp]
\centering
\caption{Top-6 word types at the positive and negative poles of PC1,
PC2 and PC3 for \emph{Ulysses} ($R=100$, $F_{\min}=5$), with
frequencies in the filtered token sequence.
$B_k$ is a property of the component, not of the pole, and is therefore
shown once per pair; the largest value over the $K=20$ components is
shown in bold.
A word type may appear at opposite poles of different components: the
components are orthogonal directions, not a partition of the
vocabulary.}
\label{tab:ulysses_pc}
\small
\setlength{\tabcolsep}{4pt}
\begin{tabular}{llr p{5.4cm} p{4.4cm}}
\toprule
PC & Pole & $B_k$ & Top words (freq) & Interpretation \\
\midrule
PC1 & $+$ & 1.299 &
  \emph{wonderworker} (6), \emph{dedalus} (174), \emph{maggy} (9),
  \emph{purefoy} (20), \emph{dignam} (83), \emph{voglio} (9)
  & Dublin proper names, neologisms and multilingual fragments
    (GloVe artefact) \\
PC1 & $-$ & &
  \emph{government} (14), \emph{any} (191), \emph{than} (169),
  \emph{people} (78), \emph{because} (219), \emph{year} (71)
  & Generic and institutional English (GloVe artefact) \\
\midrule
PC2 & $+$ & \textbf{1.470} &
  \emph{st} (6), \emph{william} (40), \emph{edward} (17),
  \emph{sir} (219), \emph{attended} (7), \emph{frederick} (10)
  & Formal titles and institutional naming \\
PC2 & $-$ & &
  \emph{yourself} (38), \emph{really} (49), \emph{feel} (104),
  \emph{stuff} (24), \emph{stupid} (11), \emph{ourselves} (10)
  & Colloquial and intimate register \\
\midrule
PC3 & $+$ & 1.324 &
  \emph{thick} (24), \emph{inches} (7), \emph{surface} (10),
  \emph{tail} (24), \emph{mud} (6), \emph{sloping} (9)
  & Physical and material description \\
PC3 & $-$ & &
  \emph{married} (56), \emph{marry} (10), \emph{poet} (29),
  \emph{sir} (219), \emph{son} (113), \emph{brother} (66)
  & Kinship, marriage and literary identity \\
\bottomrule
\end{tabular}
\end{table}

Taken together, the four texts show that PCA produces an ordered static
geometry whose leading directions are often semantically interpretable,
but whose first component can be contaminated by peripheral lexical
clusters. The temporal question is separate: whether visits along these
directions occur regularly or in concentrated textual episodes.

\FloatBarrier
\subsection{Burstiness across principal components}
\label{sec:component_burstiness_results}

We now turn from static geometry to temporal organisation. For each
component, threshold crossings define event sequences and IET
distributions. Figure~\ref{fig:burst_panel} and
Figures~\ref{fig:wrnpc_iet}, \ref{fig:darwin_iet}, \ref{fig:greatexp_iet}
and \ref{fig:ulysses_iet} show the representative-text evidence. In every text, the original
series has broader IET distributions and larger $B_k$ than either
control. The spectra are component dependent and generally largest
among the early PCs, but the decay with component rank is not monotone.
This non-monotonicity is expected because PCA orders variance rather
than burstiness.

\subsubsection{\emph{War and Peace}}

\begin{figure}[tbp]
  \centering
  \includegraphics[width=\textwidth]{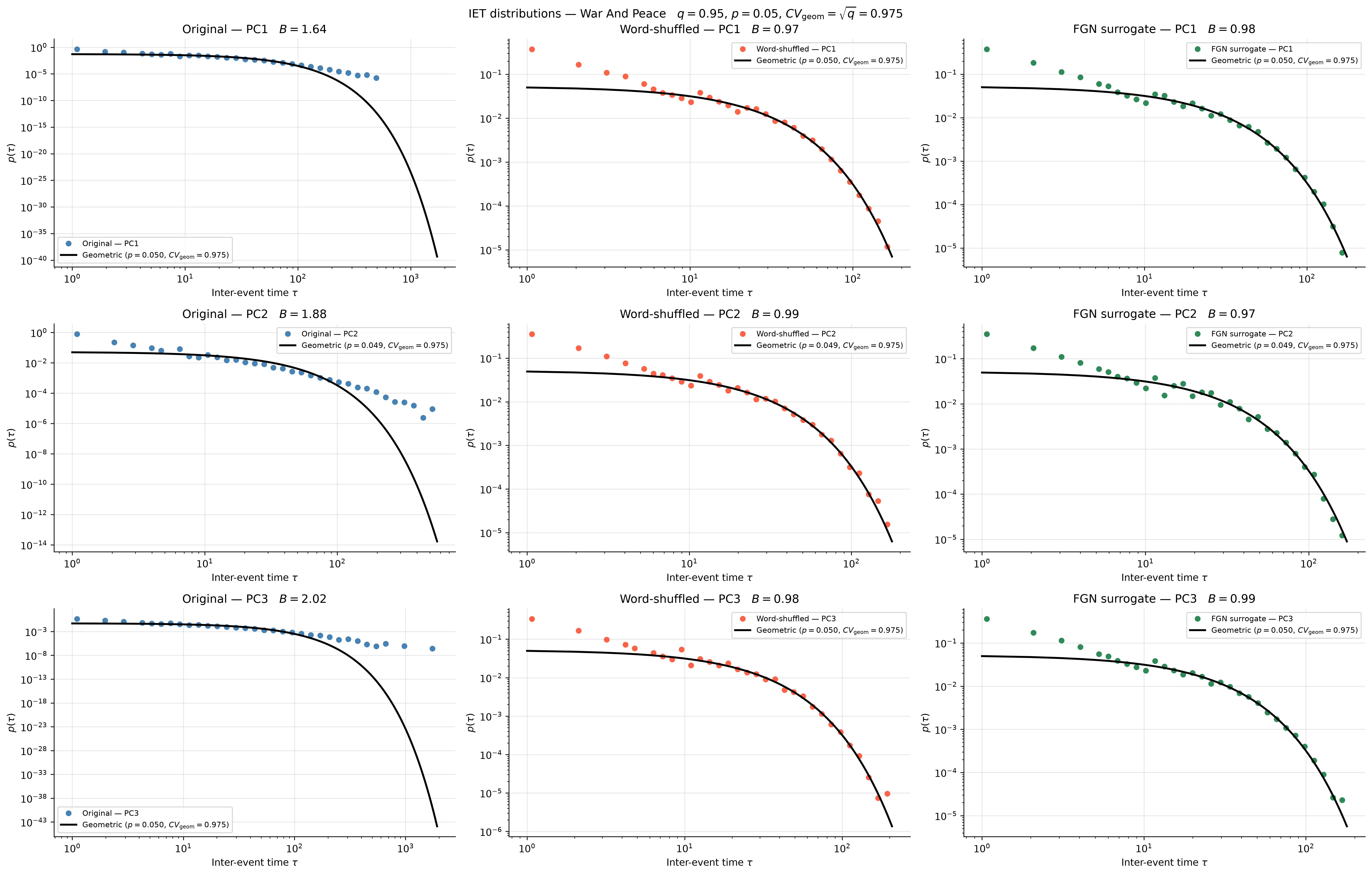}
  \caption{Inter-event-time distributions for PC1--PC3 of
  \emph{War and Peace} at $q=0.95$. Columns show the original text
  (blue), word-shuffled null (red), and FGN surrogate (green). The
  original projections have broad tails, whereas both controls lie
  close to the geometric reference.}
  \label{fig:wrnpc_iet}
\end{figure}

\begin{figure}[tbp]
  \centering
  \includegraphics[width=\textwidth]{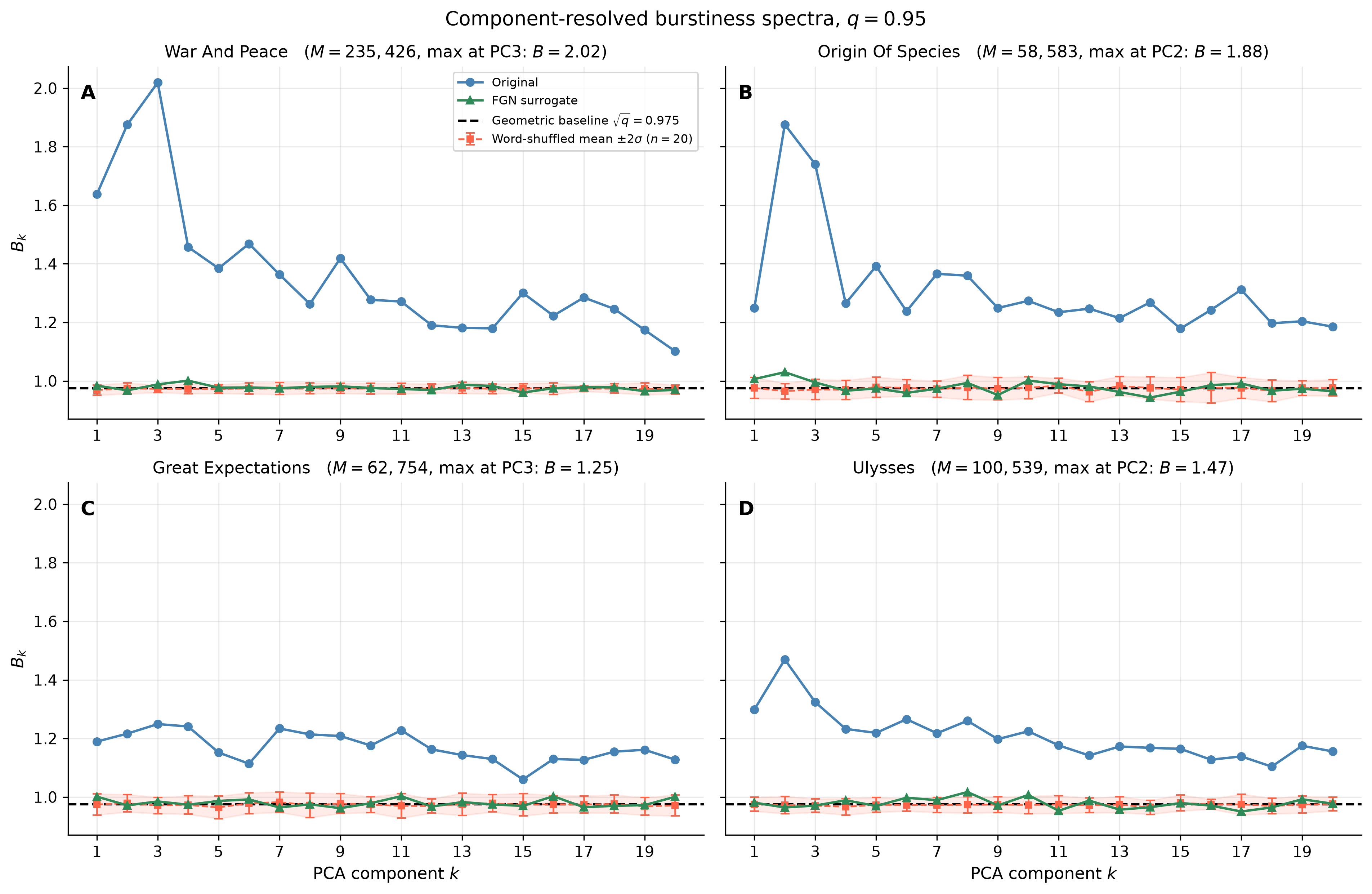}
 \caption{Component-resolved burstiness spectra at $q=0.95$ for the four
  representative texts: \textbf{A} \emph{War and Peace},
  \textbf{B} \emph{On the Origin of Species},
  \textbf{C} \emph{Great Expectations}, \textbf{D} \emph{Ulysses}.
  In every panel the original (blue circles) lies above the FGN surrogate
  (green triangles), the word-shuffled mean$\pm2\sigma$ band over 20
  realisations (red), and the geometric baseline
  $\CV_{\rm geom}=\sqrt{q}\simeq0.975$, for all 20 components.
  Events are crossings of the upper tail, so $B_k$ measures the
  clustering of visits to the \emph{positive} pole of each component
  (Tables~\ref{tab:wrnpc_pc}--\ref{tab:ulysses_pc}).
  Burstiness is not monotone in PCA rank, because PCA orders components
  by variance rather than by temporal clustering: the maximum falls at
  PC3, PC2, PC3 and PC2 respectively, and never at PC1 --- although in
  \textbf{C} the leading values are closely spaced.}
  \label{fig:burst_panel}
\end{figure}

For PC1--PC3, the original IET distributions extend well beyond the
geometric reference, whereas the shuffled and FGN controls closely
follow it (Figure~\ref{fig:wrnpc_iet}). The burstiness spectrum in
Figure~\ref{fig:burst_panel}A is above both controls for all 20
components. PC3 has the largest value, $B_3=2.019$, approximately $2.1$ times the
geometric baseline and more than twice the corresponding null values
($B_3^{\rm surr}=0.973$, $B_3^{\rm FGN}=0.988$). Because events are
threshold crossings of the upper tail, this measures the clustering of
the positive pole of PC3: passages of physical action and embodied
movement, which are concentrated in the battle and campaign chapters.
The eigenvalue--burstiness correlation is positive but imperfect,
$r=0.708$ ($P=0.0046$).

\subsubsection{\emph{On the Origin of Species}}

The original projections have broad IET tails, particularly for PC2 and
PC3, while both controls remain close to geometric
(Figure~\ref{fig:darwin_iet}). Burstiness is high across the leading
directions but peaks away from the leading eigenvalue: the largest
value is $B_2=1.876$, followed by $B_3=1.740$, while PC1 is markedly
lower at $B_1=1.250$. The positive pole of PC2 gathers places and
biogeographical setting (Table~\ref{tab:darwin_pc}), vocabulary that is
concentrated in the chapters on geographical distribution, and this is
what concentrates its threshold crossings. The full
original-over-control ordering holds for all 20 components. The linear
eigenvalue--burstiness correlation is weak and not significant,
$r=0.393$ ($P=0.089$), while the rank correlation is significant,
$\rho=0.678$ ($P=0.0015$): the association is monotone but not linear.
That the burstiness maximum does not lie at PC1 states the point
without recourse to any correlation measure --- the direction of
greatest semantic variance is not the direction of greatest temporal
clustering. Individual values $B_k$ of components whose eigenvalues are
close to their neighbours are sensitive to small rotations of the axes
(SI, Section~\ref{app:bk_stability}); PC2 is well separated from PC1
and PC3, and its value is stable.

\subsubsection{\emph{Great Expectations}}

The original PC1--PC3 IET distributions again have broader tails than
the controls (Figure~\ref{fig:greatexp_iet}). At $q=0.95$,
$B_1=1.189$, $B_2=1.217$ and $B_3=1.249$, while the null values lie in
$0.96$--$1.00$. The original remains above both controls over all 20
components (Figure~\ref{fig:burst_panel}C), but the spectrum is flatter
and lower overall than for \emph{War and Peace}: the largest value
exceeds $B_1$ by only $5\%$, against $23\%$ in Tolstoy, and the mean
over the twenty components, $1.17$, is the lowest of the four texts.
Component-level burstiness is therefore present throughout the spectrum
without being concentrated in any particular direction. The
eigenvalue--burstiness correlation is $r=0.421$ ($P=0.064$), not
significant at the $5\%$ level, whereas the rank correlation is
significant, $\rho=0.611$ ($P=0.005$).

\subsubsection{\emph{Ulysses}}

For \emph{Ulysses}, the original IET distributions are broad for all
three displayed components, while both controls collapse towards the
geometric reference (Figure~\ref{fig:ulysses_iet}). The spectrum peaks
at PC2, $B_2=1.470$, with $B_1=1.299$ and $B_3=1.324$
(Figure~\ref{fig:burst_panel}D); the positive pole of PC2 is the
register of formal titles and institutional naming
(Table~\ref{tab:ulysses_pc}), which the novel deploys in concentrated
parodic passages rather than evenly. The original exceeds both controls
throughout the first 20 components. Here too the two correlation
measures differ: $r=0.664$ ($P=0.015$) against $\rho=0.887$
($P<10^{-4}$), so that the association between semantic variance and
burstiness is close to monotone while being less well described by a
linear fit.

Averaged over the first 20 components at $q=0.95$, the representative
texts have $\bar B^{\rm orig}$ of $1.37$ (\emph{War and Peace}),
$1.32$ (\emph{On the Origin of Species}), $1.21$ (\emph{Ulysses}) and
$1.17$ (\emph{Great Expectations}), against $\bar B\simeq0.97$--$0.98$
for both null models in every case. The ordering is not a simple
function of either text length or rank-sequence DFA exponent.
The common feature is that component-level burstiness is present in the
original semantic trajectories and absent from both null models.

\FloatBarrier
\subsection{Robustness to the threshold quantile and to the number of
components}
\label{sec:single_text_robustness}

The component separation is not tied to the default $q=0.95$.
Figure~\ref{fig:sweep_panel} shows the mean
burstiness over $k=1,\ldots,20$ as $q$ varies from $0.60$ to $0.99$;
the corpus-level version of the ordering fraction is given in
Section~\ref{sec:corpus_robustness}.
For each of the four representative texts, the ordering holds for all
20 components throughout the displayed range. As expected, the
absolute burstiness increases towards more extreme thresholds because
rare visits to the upper tail accentuate episode-to-episode clustering.

The separation is likewise not an artefact of truncating the spectrum at
$K=20$. 
The retained components account for $35$--$39\%$ of the embedding
variance; refitting the PCA with $K=50$ and recomputing the spectrum at
$q=0.95$ leaves every component above the independent-event baseline in
all four texts, with a minimum of $B_k=1.037$, some $6\%$ above
$\CV_{\rm geom}$. Burstiness does decline with component index: the mean
falls from $1.17$--$1.37$ over the first twenty components to
$1.12$--$1.19$ over components $21$--$50$, while the variance carried per
component falls from $1.9\%$ to $0.65\%$. 
The choice $K=20$ is conventional rather than principled; what the
extension to $K=50$ shows is that it does not act as a threshold beyond
which component-level burstiness disappears. The leading components are
also the ones for which the semantic poles were examined
(Section~\ref{sec:semantic_geometry_results}).

\FloatBarrier
\section{Corpus-level generalisation}
\label{sec:corpus}

Having established the component-resolved pattern in four
representative texts (Section~\ref{sec:oliver}), we now assess its
generality across the full corpus.

\subsection{Corpus description}
\label{sec:corpus_texts}

The corpus comprises 28 English-language works from Project Gutenberg:
20 novels and 8 works of non-fiction (natural history, psychology,
philosophy, logic and travel writing), in English originals and two
translations, first published between 1759 and 1922
(Table~\ref{tab:corpus}). All texts are prepared with the procedure of
Section~\ref{sec:preprocessing}.
Raw token counts range from approximately $71{,}000$ (Tom Sawyer) to
$565{,}000$ (War and Peace), with filtered token counts
$M$ (after high-frequency and hapax removal, OOV excluded) between
$22{,}000$ and $235{,}000$.
DFA scaling exponents $\alpha$ lie in the range $[0.64, 0.86]$,
consistent with the literature on long-range correlations in English
texts \citep{Ebeling1995, MontemurroPury2002}.

\begin{table}[t]
\centering
\caption{Corpus of 28 texts: 20 novels (top) and 8 works of non-fiction
(bottom). Year: first publication.
$N$: total token count after tokenisation.
$M$: content tokens after frequency filtering
(top $R=100$ words removed, minimum frequency~5).
$\alpha$: DFA scaling exponent of the rank sequence.
$\alpha_0$: FGN generator exponent matched to $\alpha$ (bisection
tolerance $0.01$).}
\label{tab:corpus}
\footnotesize
\setlength{\tabcolsep}{4pt}
\begin{tabular}{llrrrll}
\toprule
Title & Author & Year & $N$ & $M$ & $\alpha$ & $\alpha_0$ \\
\midrule
Pride and Prejudice            & Austen     & 1813 & 122{,}219 &  42{,}366 & 0.642 & 0.67 \\
Jane Eyre                      & C.\ Bront\"e & 1847 & 187{,}243 &  68{,}221 & 0.682 & 0.71 \\
Wuthering Heights              & E.\ Bront\"e & 1847 & 116{,}810 &  40{,}331 & 0.721 & 0.75 \\
Don Quixote (transl.\ Ormsby)  & Cervantes  & 1605 & 404{,}384 & 147{,}764 & 0.716 & 0.76 \\
David Copperfield              & Dickens    & 1850 & 351{,}160 & 128{,}534 & 0.720 & 0.76 \\
Great Expectations             & Dickens    & 1861 & 186{,}170 &  62{,}754 & 0.703 & 0.76 \\
Oliver Twist                   & Dickens    & 1838 & 158{,}505 &  58{,}014 & 0.721 & 0.76 \\
Middlemarch                    & Eliot      & 1871 & 319{,}788 & 119{,}568 & 0.725 & 0.78 \\
Tess of the d'Urbervilles      & Hardy      & 1891 & 150{,}506 &  51{,}546 & 0.685 & 0.71 \\
Ulysses                        & Joyce      & 1922 & 263{,}007 & 100{,}539 & 0.778 & 0.80 \\
The Rainbow                    & Lawrence   & 1915 & 187{,}449 &  67{,}032 & 0.715 & 0.76 \\
Moby Dick                      & Melville   & 1851 & 211{,}867 &  80{,}978 & 0.698 & 0.73 \\
Frankenstein                   & Shelley    & 1818 &  75{,}080 &  24{,}533 & 0.709 & 0.71 \\
The Jungle                     & Sinclair   & 1906 & 151{,}114 &  49{,}524 & 0.773 & 0.83 \\
Dracula                        & Stoker     & 1897 & 161{,}221 &  55{,}315 & 0.768 & 0.83 \\
Vanity Fair                    & Thackeray  & 1848 & 305{,}028 & 120{,}171 & 0.717 & 0.76 \\
War and Peace (transl.\ Maude) & Tolstoy    & 1869 & 564{,}609 & 235{,}426 & 0.724 & 0.76 \\
Huckleberry Finn               & Twain      & 1884 & 111{,}380 &  34{,}640 & 0.736 & 0.78 \\
Tom Sawyer                     & Twain      & 1876 &  70{,}817 &  21{,}855 & 0.792 & 0.80 \\
The Picture of Dorian Gray     & Wilde      & 1891 &  79{,}257 &  25{,}728 & 0.844 & 0.87 \\
\midrule
On the Origin of Species       & Darwin     & 1859 & 150{,}812 &  58{,}583 & 0.772 & 0.80 \\
The Voyage of the Beagle       & Darwin     & 1839 & 196{,}103 &  79{,}752 & 0.646 & 0.69 \\
Psychology: Briefer Course     & James      & 1892 & 150{,}453 &  55{,}898 & 0.695 & 0.73 \\
A System of Logic              & Mill       & 1843 & 371{,}981 & 142{,}532 & 0.772 & 0.83 \\
The Analysis of Mind           & Russell    & 1921 &  88{,}221 &  30{,}515 & 0.752 & 0.76 \\
The Theory of Moral Sentiments & Smith      & 1759 & 105{,}956 &  35{,}602 & 0.713 & 0.73 \\
Life on the Mississippi        & Twain      & 1883 & 130{,}313 &  45{,}856 & 0.742 & 0.78 \\
Island Life                    & Wallace    & 1880 & 130{,}882 &  50{,}061 & 0.861 & 0.92 \\
\bottomrule
\end{tabular}
\end{table}

\begin{quote}
The corpus is rebuilt from the Project Gutenberg files by
\texttt{scripts/\allowbreak build\_corpus.py}, and per-text IET
distributions and burstiness spectra for all texts and threshold values
can be regenerated with \texttt{scripts/\allowbreak run\_text.py} in
the code repository.
The main text reports the aggregated corpus results.
\end{quote}

\subsection{Component-resolved burstiness holds corpus-wide}
\label{sec:corpus_ordering}

\begin{figure}[tbp]
  \centering
  \includegraphics[width=\textwidth]{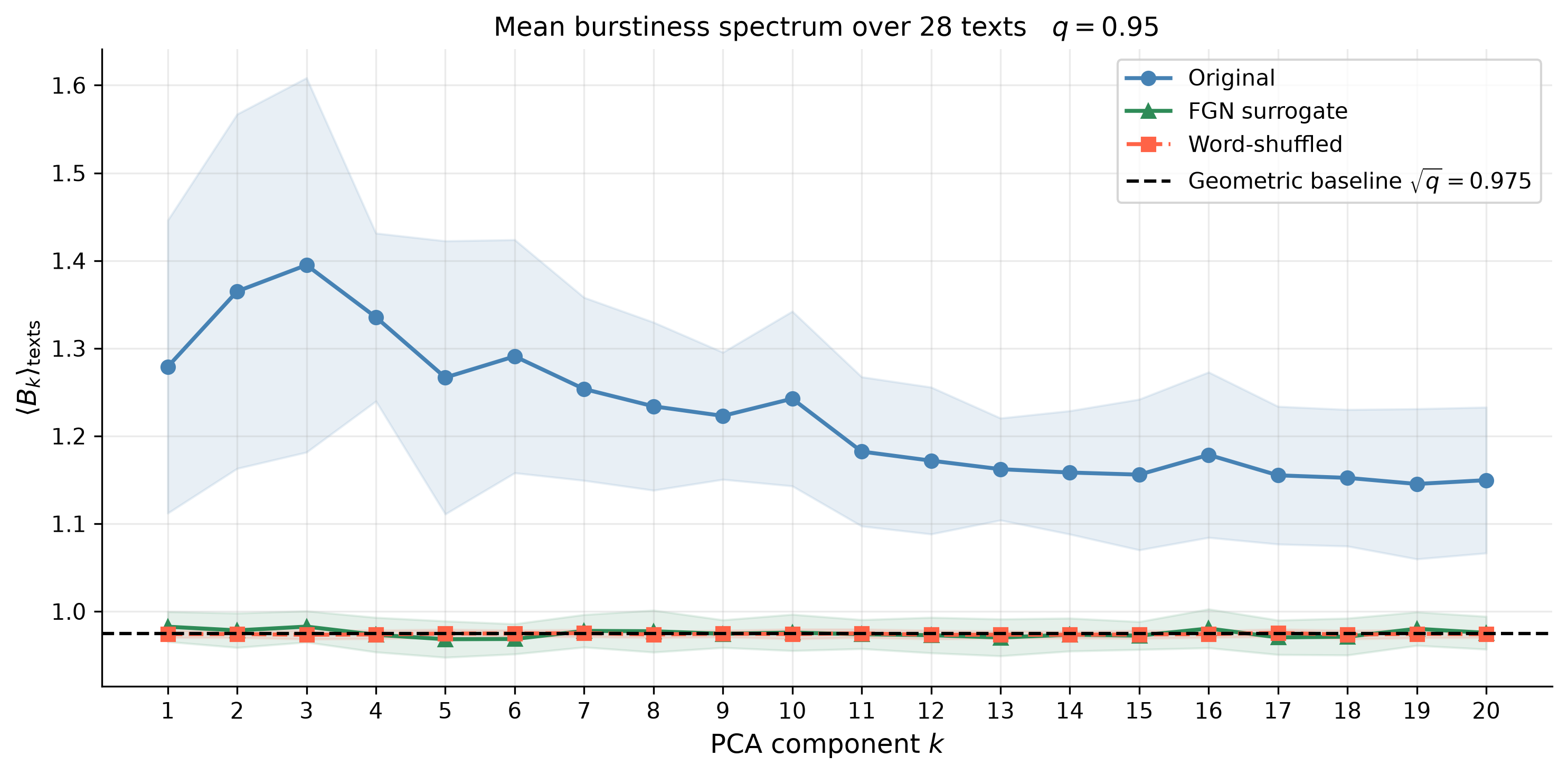}
  \caption{Mean component-resolved burstiness spectrum across the
  analysed corpus at $q=0.95$. Curves show the mean $\langle B_k\rangle$
  over texts for $k=1,\ldots,20$; shaded bands indicate one standard
  deviation. The original texts (blue) are above the FGN surrogate
  (green), word-shuffled null (red), and geometric baseline
  $\sqrt{q}\simeq0.975$ at every component. The largest corpus means
  occur among the early components, followed by a broad decline with
  non-monotone secondary peaks.}
  \label{fig:corpus_spectrum}
\end{figure}

The corpus-level component spectrum is summarised in
Figure~\ref{fig:corpus_spectrum}.
Averaging the burstiness spectrum $\{\bk\}$ across all 28 texts
at quantile $q = 0.95$, the original texts are systematically more
bursty than both null models across all 20 PCA components:
$\langle B_k^{\rm orig}\rangle_{\rm texts} \in [1.15, 1.39]$
(mean $1.22$), while both null models cluster tightly around
the geometric baseline $\CV_{\rm geom} = \sqrt{0.95} \approx 0.975$:
$\langle B_k^{\rm surr}\rangle \in [0.973, 0.975]$ and
$\langle B_k^{\rm FGN}\rangle \in [0.968, 0.983]$.

At the level of individual $({\rm text}, k)$ pairs
($28 \times 20 = 560$ pairs):
\begin{itemize}
  \item $B_k^{\rm orig} > B_k^{\rm surr}$ in $\mathbf{100\%}$ of pairs.
  \item $B_k^{\rm orig} > B_k^{\rm FGN}$ in $\mathbf{100\%}$ of pairs.
  \item $|B_k^{\rm FGN}/B_k^{\rm surr} - 1| < 0.05$ in
        $\mathbf{97.7\%}$ of pairs.
\end{itemize}

The appropriate summary statistic for the magnitude of the effect
is the \emph{per-text mean ratio}, computed by averaging
$B_k^{\rm orig}/B_k^{\rm surr}$ over the 20 components for each
text and then averaging over the 28 texts:
\begin{equation}
  \left\langle \frac{B_k^{\rm orig}}{B_k^{\rm surr}}
  \right\rangle_{\rm texts}
  \;=\; 1.26 \;\pm\; 0.01,
  \qquad t = 18.1,\quad P < 10^{-15},
  \label{eq:ratio_corpus}
\end{equation}
where $\pm$ denotes the standard error of the mean over 28 texts.
The original texts are thus on average $\sim 26\%$ more bursty
than the word-shuffled null, which itself sits at the geometric baseline.
 The per-text values range from $1.15$ (\emph{Wuthering Heights}) to
$1.44$ (\emph{Island Life}), with a spread across texts
(sd $= 0.075$) that is not explained by text length
($r=0.22$, $P=0.26$, against $M$). Part of the spread is associated
with genre: the eight non-fiction works have a mean ratio of
$1.331 \pm 0.021$, against $1.228 \pm 0.013$ for the twenty novels
(Welch $t=4.1$, $P=0.001$; Mann--Whitney $P=2\times10^{-4}$), and seven
of them occupy the eight highest positions. This is consistent with the
chapter-level topical segmentation of expository prose. With only eight
non-fiction texts, chosen as canonical works rather than sampled, the
difference is a clear feature of this corpus but should not be read as
an estimate for expository prose in general.
The analogous ratio for the FGN surrogate is
$\langle B_k^{\rm orig}/B_k^{\rm FGN}\rangle = 1.26 \pm 0.01$,
indistinguishable from the word-shuffled value.

The two null models satisfy
$\langle B_k^{\rm FGN}/B_k^{\rm surr}\rangle = 1.001 \pm 0.001$
(range over texts $[0.989, 1.006]$; $t = 1.2$, $P = 0.23$):
the FGN surrogate and the word-shuffled null are statistically
indistinguishable at the corpus level
(Section~\ref{sec:corpus_diffs}).

\subsection{Robustness to the threshold quantile}
\label{sec:corpus_robustness}

\begin{figure}[tbp]
  \centering
  \includegraphics[width=0.72\textwidth]{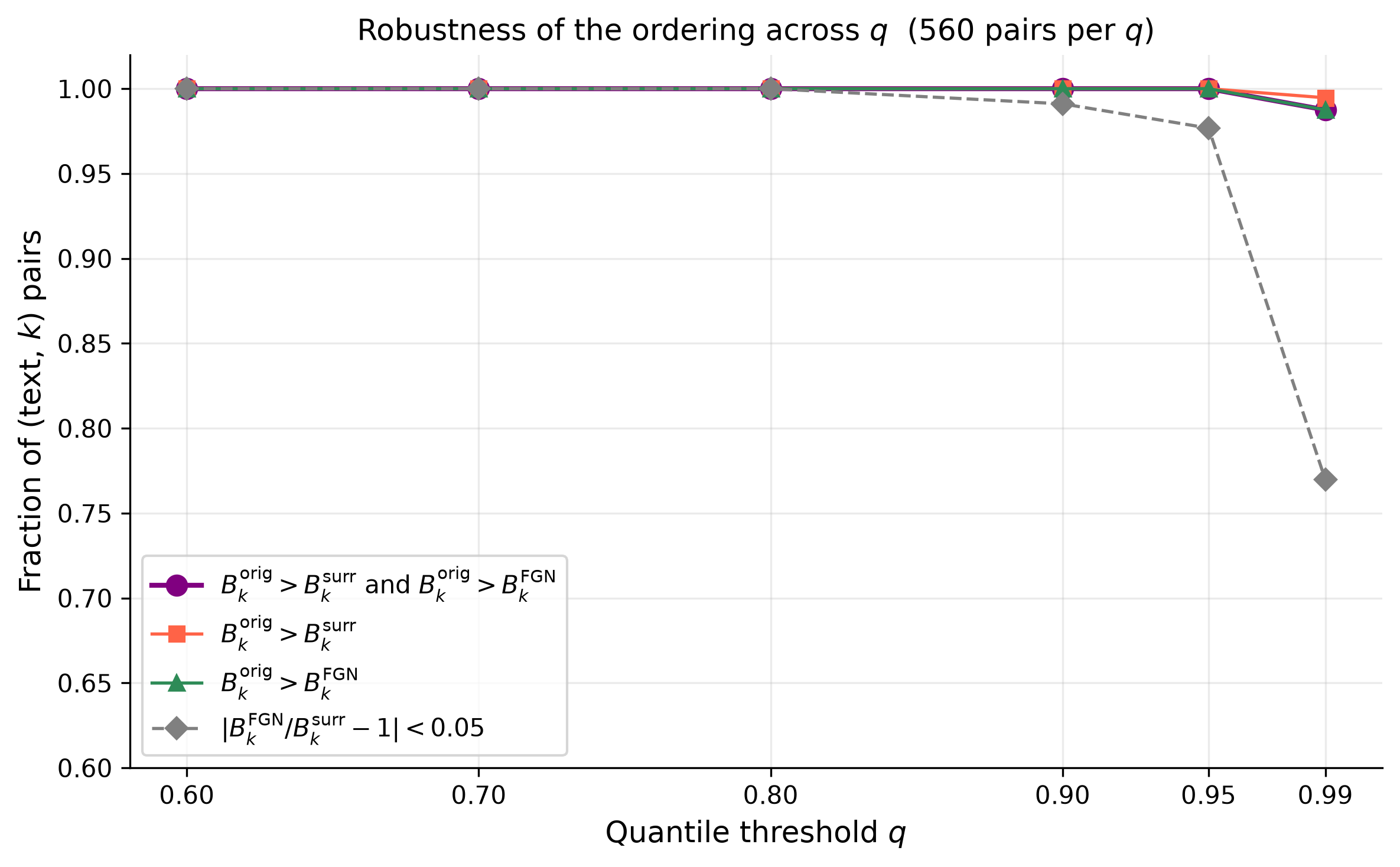}
  \caption{%
    \textbf{Robustness of the burstiness ordering across quantile
    threshold $q$.}
    Each point: fraction of $({\rm text}, k)$ pairs ($560$ per $q$)
    satisfying the indicated condition.
    Purple circles: full ordering
    $B_k^{\rm orig} > B_k^{\rm surr}$ and $B_k^{\rm orig} >
    B_k^{\rm FGN}$.
    Red squares: $B_k^{\rm orig} > B_k^{\rm surr}$.
    Green triangles: $B_k^{\rm orig} > B_k^{\rm FGN}$.
    Grey diamonds: $|B_k^{\rm FGN}/B_k^{\rm surr} - 1| < 0.05$.
    The ordering is robust for $q \in [0.60, 0.95]$; the slight
    degradation at $q = 0.99$ is due to estimator noise at
    $p = 0.01$.
  }
  \label{fig:corpus_robustness}
\end{figure}

Figure~\ref{fig:corpus_robustness} shows the fraction of
$({\rm text}, k)$ pairs satisfying the ordering conditions as a
function of $q \in \{0.60, 0.70, 0.80, 0.90, 0.95, 0.99\}$.

The full ordering $B_k^{\rm orig} > B_k^{\rm surr}$ and
$B_k^{\rm orig} > B_k^{\rm FGN}$ is satisfied in $100\%$
of pairs for all $q \in [0.60, 0.95]$, and in $98.8\%$ of pairs at the
extreme $q = 0.99$.
The null-model indistinguishability criterion
$|B_k^{\rm FGN}/B_k^{\rm surr} - 1| < 0.05$ is met by $\geq 99\%$ of
pairs for $q \leq 0.90$, by $98\%$ at $q = 0.95$, and decreases to
$77\%$ at $q = 0.99$.

This degradation at the extreme threshold is expected: at the corpus
median, $M \approx 57{,}000$, the event rate $p = 0.01$ leaves only
$\sim 570$ events per series, giving a relative error on the IET
estimator of $\sim 4\%$, rising to $\sim 7\%$ for the shortest text
($M = 21{,}855$) --- comparable to or larger than the magnitude of
$B^{\rm FGN} - B^{\rm surr}$ at this threshold.
For $q = 0.95$ the same texts give $\sim 2{,}800$ and $\sim 1{,}100$ events,
with relative errors of $1.9\%$ and $3.0\%$, well within the regime
where the null-model comparison is reliable.

The same estimator noise accounts for the seven pairs out of 560 that fail the full ordering at $q=0.99$; five of them belong to \emph{Wuthering Heights}, the text with the lowest mean ratio.

\subsection{Distribution of burstiness ratios}
\label{sec:corpus_diffs}

Figure~\ref{fig:corpus_diffs} shows the distribution of the ratios
$B^{\rm orig}/B^{\rm surr}$, $B^{\rm orig}/B^{\rm FGN}$, and
$B^{\rm FGN}/B^{\rm surr}$ over all 560 $({\rm text}, k)$ pairs
at $q = 0.95$.

Panel~A shows that both orig--null ratios are concentrate
well above 1, with no mass below 1.
The two distributions (orig/surr and orig/FGN) are nearly
identical in shape and location, confirming that the FGN and
word-shuffled null models are equally poor surrogates for the
original burstiness.

Panel~B isolates the ratio $B^{\rm FGN}/B^{\rm surr}$, which is tightly
concentrated around 1 over the 560 pairs (mean $1.0009$, sd $0.021$,
range $[0.93, 1.11]$). Because components within a text are not
independent, the test is performed on the 28 per-text means: a
one-sample $t$-test against unity gives $t = 1.22$, $P = 0.23$. The two
null models are statistically indistinguishable, and the distribution
is symmetric with no indication of a directional bias.

\paragraph{FGN surrogate matched after filtering.}
Matching the rank-sequence exponent before filtering leaves the retained
tokens of the FGN surrogate with a weaker rank-level memory than those of
the original in 22 of the 28 texts (Section~\ref{sec:null_model_fgn}).
This loss does not account for the absence of burstiness in the control.
Across texts, $\langle B_k^{\rm FGN}/B_k^{\rm surr}\rangle$ is
uncorrelated with the difference between the filtered FGN and original
exponents ($r=0.00$), and in the six texts in which the filtered FGN
sequence has the larger exponent --- by up to $0.09$ in \emph{Island
Life} --- the ratio stays between $0.995$ and $1.003$.
The surrogate matched after filtering, FGN$_M$, reproduces the exponent
of the original retained tokens to within $0.01$ in every text. Its
burstiness ratio,
$\langle B_k^{{\rm FGN}_M}/B_k^{\rm surr}\rangle = 1.006 \pm 0.001$, is
statistically above unity ($t=5.3$) but amounts to $2.4\%$ of the excess
of the original texts ($1.257$), and the full ordering
$B_k^{\rm orig}>B_k^{{\rm FGN}_M}$, $B_k^{\rm orig}>B_k^{\rm surr}$ holds
in all 560 pairs at $q=0.95$ and in at least $99\%$ of the pairs at every
$q$. The small excess is concentrated on PC1 and PC3 (mean
$B_k^{{\rm FGN}_M}/B_k^{\rm surr}=1.022$ on both), which are also the
components on which this surrogate shows significant lag-one
autocorrelation in most texts (100\% and 57\% of the texts, against
13\% for $k\geq4$); for $k\geq4$ the excess is $1.9\%$ of that of the
original. The mean projection DFA exponent of FGN$_M$ is $0.510$,
against $0.663$ for the originals and $0.504$ for the standard FGN
surrogate (SI, Section~\ref{app:filtered_fgn}). Rank-level scaling
matched on the retained content words therefore transmits a detectable
but very small amount of clustering to semantic projections, mostly
along PC1, the direction associated with word frequency, and PC3, and
essentially none along the remaining components.

Together, Figures~\ref{fig:corpus_spectrum}--\ref{fig:corpus_diffs}
establish the corpus-level result:
\begin{equation}
  B_k^{\rm orig} \;\gg\; B_k^{\rm FGN} \;\approx\; B_k^{\rm surr}
  \qquad \forall\, k,\; \forall\, \text{text in corpus},
  \label{eq:ordering_corpus}
\end{equation}
with a mean excess of the original over the word-shuffled null of
$\sim 26\%$. The ordering holds for every text and every component
irrespective of length, genre, DFA exponent and choice of quantile
threshold; its magnitude, by contrast, varies systematically between
texts and is larger for non-fiction (Section~\ref{sec:corpus_ordering}).

\FloatBarrier
\section{Temporal persistence in semantic projection space}
\label{sec:temporal_persistence}

Sections~\ref{sec:oliver} and~\ref{sec:corpus} establish that the
original semantic trajectories have elevated component-resolved
burstiness and are separated from both controls. We now test temporal
persistence directly in the continuous semantic projections using
normalised autocorrelation functions and DFA.

\subsection{Autocorrelation of projections}
\label{sec:autocorr_proxy}

For PCA component $k$ and lag $\ell$, we compute the normalised
autocorrelation function
\begin{equation}
  \rho_k(\ell) \;=\;
  \frac{\mathrm{Cov}\!\left(\xtk,\, x_{t+\ell}^{(k)}\right)}
       {\mathrm{Var}\!\left(\xtk\right)},
  \label{eq:autocorr}
\end{equation}
which is zero in expectation for an i.i.d. sequence. The word-shuffled
null therefore provides a direct finite-sample reference. For the FGN
surrogate, rank-level memory does not by itself imply autocorrelation
along a semantic projection because rank proximity is not semantic
proximity (Remark~\ref{rem:rank_semantic}); whether such persistence
survives the rank-to-embedding mapping is an empirical question.

\begin{remark}
  The ACF is connected to the burstiness spectrum through the
  Wiener--Khintchine theorem: a slowly decaying $\rho_k(\ell)$
  corresponds to excess power at low frequencies in the spectral
  density of $x_t^{(k)}$. For a Gaussian process the level-crossing
  statistics are determined by the autocorrelation; in general the two
  are related but not equivalent, since $\bk$ depends on the full IET
  distribution while $\rho_k$ is a second-order statistic. In the
  present data they move together: the original has both elevated
  $\bk$ and slowly decaying $\rho_k$, and both null models have
  neither.
\end{remark}

\subsection{Empirical results: ACF of projections}
\label{sec:acf_results}

\begin{figure}[tbp]
  \centering
  \includegraphics[width=\textwidth]{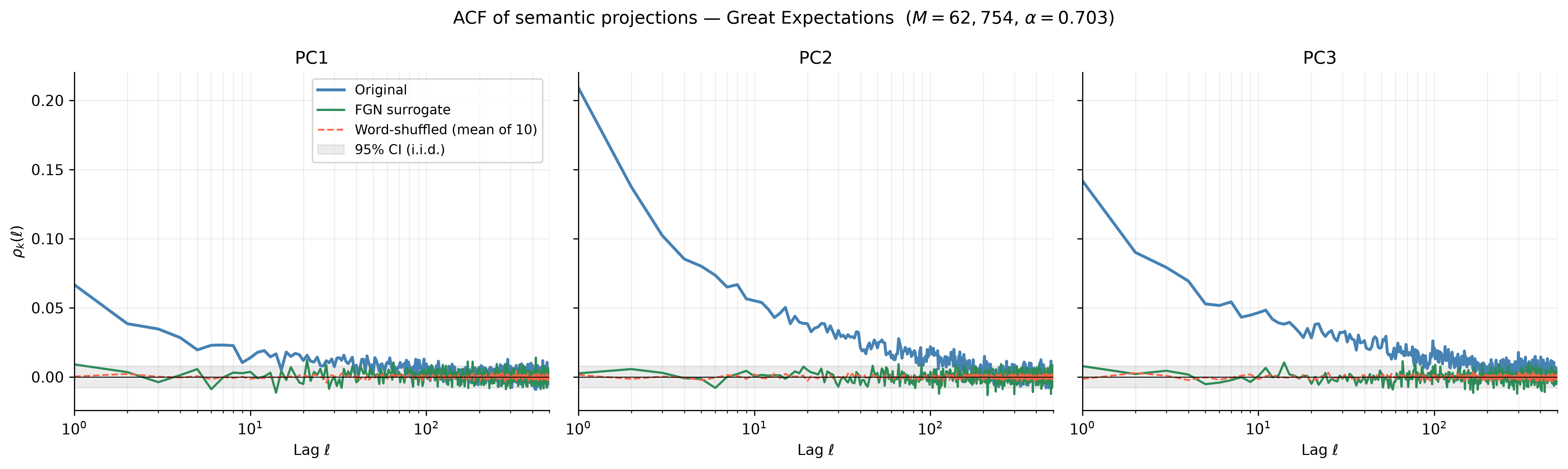}
  \caption{%
    \textbf{Autocorrelation function $\rho_k(\ell)$ of projected
    series, as a function of lag $\ell$ (log scale) ---
    \emph{Great Expectations} (Dickens).}
    Each panel shows one PCA component (PC1, PC2, PC3).
    Blue: original filtered text.
    Green: FGN surrogate.
    Red dashed: word-shuffled null.
    Grey band: $\pm 1.96/\sqrt{M}$ i.i.d.\ reference (95\% CI).
    The original text has substantial positive ACF that decays slowly
    on a log scale, remaining above the i.i.d.\ band up to
    $\ell \sim 100$.
    Both null models collapse immediately into the i.i.d.\ band,
    despite the FGN surrogate having the same DFA exponent
    $\alpha = 0.703$ as the original.
  }
  \label{fig:acf_decay}
\end{figure}

Figure~\ref{fig:acf_decay} shows $\rho_k(\ell)$ as a function of lag
$\ell$ (log scale) for PC1, PC2, and PC3 of
\emph{Great Expectations} by Dickens ($\alpha = 0.703$), for the original text, the FGN surrogate, and
the word-shuffled null.
Figure~\ref{fig:acf_mean} shows the same for the mean ACF over all
$K = 20$ components.

\paragraph{Original text.}
The ACF is positive and significantly above the i.i.d.\ band
$\pm 1.96/\sqrt{M} \approx \pm 0.008$ at lag $\ell = 1$ for all 20
components, with $\rho_k(1) \in [0.050, 0.209]$
(Table~\ref{tab:acf_summary}).
The decay is gradual on a log scale, with values remaining above the
i.i.d.\ band up to lags of order $\ell \sim 100$ for the
lower-indexed components.
This slow decay is a second-order signature of the temporal
persistence that accompanies semantic burstiness.

\paragraph{FGN surrogate.}
All 20 components have $|\rho_k^{\rm FGN}(1)| \le 0.009$, and 18 of
them lie inside the i.i.d.\ band; at lag $\ell = 100$, again 18 of 20
lie inside it ($|\rho_k^{\rm FGN}(100)| \le 0.012$), the number expected
from the $5\%$ false-positive rate of the band.
The FGN surrogate, despite matching the DFA exponent of the original
($\alpha = 0.703$), exhibits only negligible autocorrelation in the
projected series: rank-level memory does not reach the thematic
directions of semantic space.

\paragraph{Word-shuffled null.}
As expected, $\rho_k^{\rm surr}(\ell) \approx 0$ for all $k$ and
$\ell$, consistent with the random-permutation reference.

\subsection{Corpus-level ACF results}
\label{sec:acf_corpus}

We computed $\rho_k(\ell)$ for all 28 texts and all $K=20$ PCA
components at lags $\ell \in \{1, 2, 5, 10, 20, 50, 100, 200, 300\}$,
yielding $28 \times 20 \times 9 = 5{,}040$ measurements.
The 95\% confidence band for an i.i.d.\ process at sample size $M$
is $\pm 1.96/\sqrt{M}$; since $M$ varies across texts, we use the
text-specific CI for significance tests.

\paragraph{Mean ACF decay.}
Figure~\ref{fig:acf_corpus_decay} shows the mean $\rho(\ell)$ averaged
over all $({\rm text}, k)$ pairs as a function of lag.
The original text maintains positive and significant autocorrelation
across the full range of lags tested, with mean values
\begin{equation}
  \bar\rho^{\rm orig}(1) = 0.099 \pm 0.042,
  \quad
  \bar\rho^{\rm orig}(10) = 0.030 \pm 0.019,
  \quad
  \bar\rho^{\rm orig}(100) = 0.011 \pm 0.011,
  \label{eq:rho_orig_values}
\end{equation}
where $\pm$ denotes one standard deviation over $({\rm text}, k)$
pairs.
The FGN surrogate and the word-shuffled null are indistinguishable
from each other and from zero across all lags:
\begin{equation}
  \bar\rho^{\rm FGN}(1) = 0.001 \pm 0.005,
  \qquad
  \bar\rho^{\rm surr}(1) = 0.000 \pm 0.004.
  \label{eq:rho_null_values}
\end{equation}
The FGN rank-to-embedding mapping thus removes essentially all of the
lag-one autocorrelation present in the original semantic projections,
even though the surrogate retains the long-range correlations of the
rank sequence.

\paragraph{Fraction of significant pairs.}
Figure~\ref{fig:acf_corpus_frac} shows, for each lag, the fraction of
$({\rm text}, k)$ pairs where $|\rho_k(\ell)|$ exceeds the i.i.d.\
band.
At lag $\ell = 1$: $100\%$ of original-text pairs are significant,
$9.5\%$ of FGN pairs, and $3.8\%$ of shuffled pairs.
The FGN figure exceeds the expected false-positive rate
of $5\%$, but the significant cases are concentrated on PC1
(68\% of the PC1 pairs, against 6\% for $k \geq 4$) --- the component
most affected by the geometric artefact of rare transliterated proper
nouns in the GloVe embedding space (Section~\ref{sec:filtering_results}).
From lag $\ell = 2$ onwards the FGN fraction lies between $3.8\%$ and
$5.9\%$, at the false-positive level. The effect is in any case tiny
relative to the separation of the original projections from both
controls.

The ``original only'' curve (orig significant, FGN not) peaks at
$94\%$ at lag $\ell = 2$ and remains above $50\%$ up to lag
$\ell = 100$, providing a clean corpus-level version of the
single-text result.
The original text maintains detectable semantic autocorrelation
at scales of order $100$ filtered tokens, corresponding to roughly $300$ tokens 
of running text --- a few paragraphs.

\paragraph{Distribution of $\rho_k(1)$.}
Figure~\ref{fig:acf_corpus_rho1} shows the distribution of
$\rho_k(1)$ over all $560$ $({\rm text}, k)$ pairs.
Panel~A shows two completely separated distributions: the original
has $\rho_k(1) \in [0.027, 0.249]$ with mean $0.099$, all
pairs significant; the FGN has $\rho_k(1) \in [-0.014, 0.027]$
with mean $0.001$, the vast majority inside the i.i.d.\ band.

Panel~B zooms on the FGN and shuffled distributions.
Both are centred near zero and overlap substantially.
A one-sample $t$-test of $\bar\rho^{\rm FGN}(1) = 0$ gives
$t = 5.94$, $P < 10^{-8}$: the mean is statistically non-zero but
negligibly small ($0.001$) relative to the orig--null gap ($0.097$).
As with the burstiness differences (Section~\ref{sec:corpus_diffs}),
the statistical significance at this sample size reflects estimator
sensitivity rather than a scientifically meaningful effect.

\begin{remark}
  The residual FGN autocorrelation on PC1 is expected rather than
  anomalous. Rank is an ordering by frequency, so the long-range memory
  imposed on $\mathbf{r}$ is memory along whichever direction of
  embedding space encodes frequency --- and in every text of the corpus
  that direction is PC1, whose poles oppose rare proper nouns and
  transliterations to generic high-frequency vocabulary
  (Section~\ref{sec:semantic_geometry_results}).
  The observation that $68\%$ of PC1 pairs but only $6\%$ of pairs with
  $k \geq 4$ show significant lag-one autocorrelation is therefore the
  signature of rank-level memory reaching the one direction it can
  reach, and failing to reach the thematic ones.
\end{remark}

\subsection{DFA of projected series}
\label{sec:dfa_projections}

The ACF results of Sections~\ref{sec:acf_results}
and~\ref{sec:acf_corpus} confirm the predicted ordering
$\rho_k^{\rm orig}(\ell) \gg \rho_k^{\rm FGN}(\ell) \approx
\rho_k^{\rm surr}(\ell) \approx 0$.
We now complement the ACF analysis by computing the DFA scaling
exponent of the projected series $x_t^{(k)}$ directly, providing the
semantic-level analogue of the
symbolic-level DFA result that motivates the paper.

For each PCA component $k$ and each of the three series
(original, FGN, word-shuffled), we apply DFA to the scalar
time series $x_t^{(k)}$ and estimate the scaling exponent
$\alpha_k$ from $F(s) \sim s^{\alpha_k}$, where $F(s)$ is the
detrended fluctuation function at box size $s$.
Box sizes range from $s_{\min} = 10$ to $s_{\max} = M/4$
on a logarithmic grid.

\begin{figure}[tbp]
  \centering
  \includegraphics[width=\textwidth]{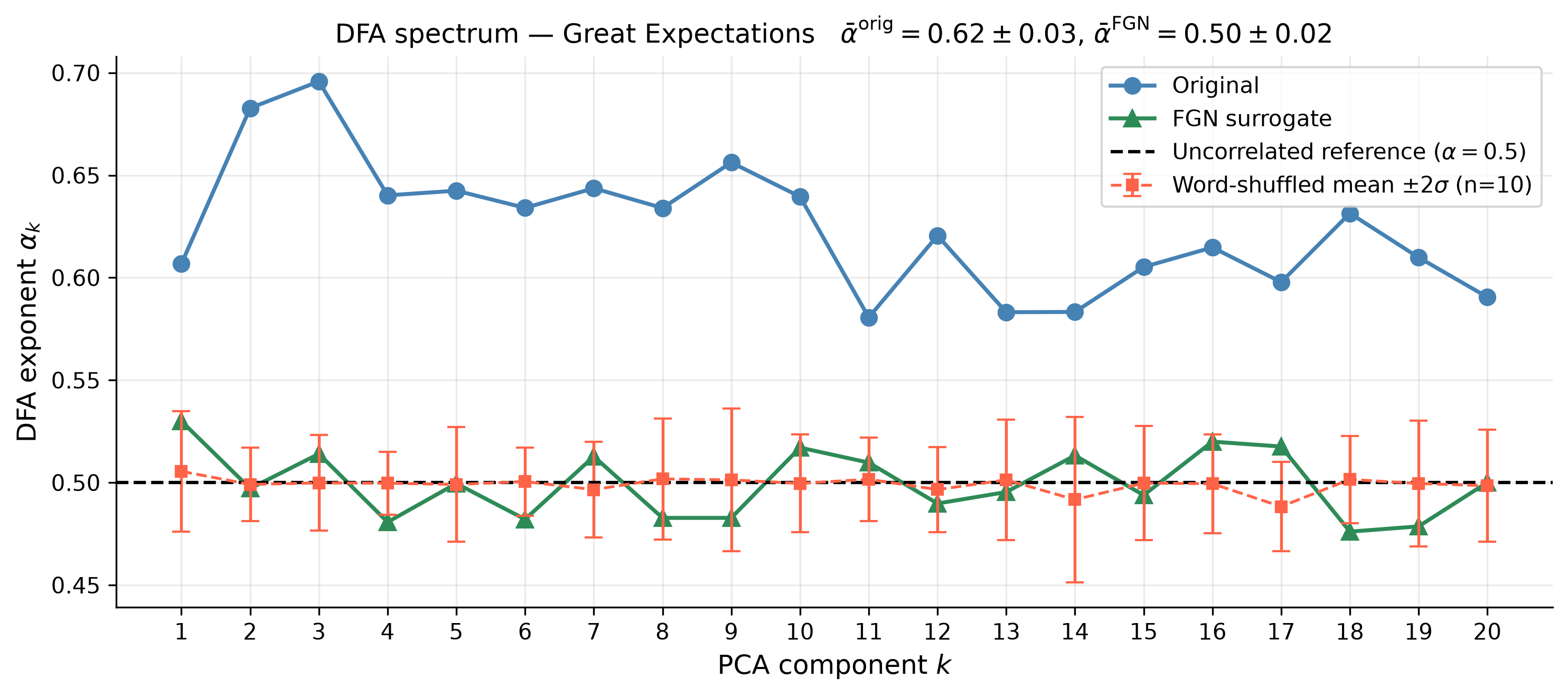}
  \caption{%
    \textbf{DFA scaling exponent $\alpha_k$ of projected series
    $x_t^{(k)}$ as a function of PCA component $k$ ---
    \emph{Great Expectations} (Dickens).}
    Blue circles: original filtered text.
    Green triangles: FGN surrogate.
    Red squares: word-shuffled null (mean $\pm 2\sigma$
    over 10 realisations).
    Dashed line: uncorrelated reference $\alpha = 0.5$.
    The original text has $\alpha_k > 0.5$ for all 20 components;
    the FGN and shuffled are indistinguishable from $\alpha = 0.5$.
  }
  \label{fig:dfa_spectrum}
\end{figure}

\paragraph{Results on Great Expectations.}
Table~\ref{tab:dfa_proj} and Figure~\ref{fig:dfa_spectrum}
report $\alpha_k$ for all $K = 20$ components of
\emph{Great Expectations} by Dickens ($\alpha = 0.703$).
The results confirm the prediction precisely:

\begin{itemize}
  \item \textbf{Original}: $\alpha_k^{\rm orig} > 0.5$ in all
        $20/20$ components, with mean
        $\bar\alpha^{\rm orig} = 0.625 \pm 0.032$.
        All values exceed the shuffled reference by more than
        $2\sigma$, confirming long-range correlations in every
        semantic projection.

  \item \textbf{FGN surrogate}: $|\alpha_k^{\rm FGN} - 0.5| < 0.05$
        in all $20/20$ components, with mean
        $\bar\alpha^{\rm FGN} = 0.500 \pm 0.016$.
        Despite having a matched DFA exponent ($\alpha=0.703$,
        generator exponent $\alpha_0=0.76$) in the rank sequence
        by construction, the FGN projects
        onto every semantic direction as an uncorrelated process.

  \item \textbf{Word-shuffled}: $\bar\alpha^{\rm surr} =
        0.499 \pm 0.004$, consistent with $\alpha = 0.5$
        (white noise) as expected for an i.i.d.\ sequence.
\end{itemize}

Figure~\ref{fig:dfa_loglog} in the Supplementary Material shows the
log-log fluctuation function $F(s)$ for PC1, PC2, and PC3.
The original text shows a clear power-law scaling with
slope $\alpha_k \approx 0.61$--$0.70$, visibly steeper than
the reference slope $0.5$.
The FGN and shuffled curves are parallel to the reference slope,
confirming $\alpha_k \approx 0.5$ for both null models.

\begin{remark}
  The exponents $\alpha_k^{\rm orig} \in [0.581, 0.696]$ do not
  decrease monotonically with $k$: PC9 ($\alpha=0.656$) and PC12
  ($\alpha=0.621$) exceed PC1 ($\alpha=0.607$), even though they carry
  several times less variance.
  This partial decorrelation between PCA rank (ordered by variance) and
  DFA exponent mirrors the decorrelation between the eigenvalue
  spectrum $\{\lambda_k\}$ and the burstiness spectrum $\{B_k\}$
  reported in Section~\ref{sec:oliver}: semantic variance and semantic
  temporal structure are distinct properties of narrative
  organisation.
\end{remark}

\paragraph{Consistent temporal signatures.}
The DFA, ACF, and burstiness analyses provide mutually consistent but
non-equivalent summaries of temporal organisation. For the original
projections, $\alpha_k>0.5$, the ACF remains positive over extended
lags, and $B_k$ exceeds the geometric reference. For both controls,
$\alpha_k\simeq0.5$, the ACF is close to zero, and
$B_k\simeq\CV_{\rm geom}$. These observations can be summarised as
\begin{equation}
  \alpha_k^{\rm orig}>0.5,
  \qquad \rho_k^{\rm orig}(\ell)>0\ \text{over extended lags},
  \qquad B_k^{\rm orig}>\CV_{\rm geom},
  \label{eq:chain_numbers}
\end{equation}
whereas the FGN and shuffled projections remain close to their
uncorrelated references in all three measures.

The FGN surrogate is especially informative because its rank sequence
has the prescribed long-range DFA exponent, while every analysed
semantic projection has $\alpha_k^{\rm FGN}\simeq0.5$. Rank-level
persistence therefore does not propagate automatically into semantic
projection space.

\subsection{Connection between ACF and burstiness}
\label{sec:acf_burstiness_connection}

The ACF and the burstiness parameter probe different aspects of the
same projected series. By the Wiener--Khintchine theorem, slow ACF
decay is associated with excess low-frequency power in the spectral
density
\begin{equation}
  S_k(\omega)=\sum_{\ell=-\infty}^{\infty}
  \rho_k(\ell)e^{-i\omega\ell}.
  \label{eq:projection_spectrum}
\end{equation}
Such low-frequency variation is consistent with extended semantic
episodes and clustered threshold exceedances, and therefore with broad
IET distributions; in the long-range correlated case the sum
in~\eqref{eq:projection_spectrum} diverges as $\omega\to0$, which is
the spectral expression of that persistence. The connection is not one-to-one, however:
thresholding is non-linear, and $B_k$ depends on temporal structure
beyond the second-order information contained in the ACF.

The empirical conclusion is consequently based on the agreement of
three complementary observations rather than on a deterministic causal
identity. Original texts show elevated $B_k$, positive projection ACFs,
and projection DFA exponents above $0.5$. The FGN and word-shuffled controls remain 
close to the geometric
baseline in burstiness, to zero in autocorrelation, and to white noise
in DFA. This convergence of measures supports the
interpretation that ordered topical organisation, rather than global
rank-level scaling alone, produces persistence in semantic projection
space.

\FloatBarrier
\section{Discussion and Conclusions}
\label{sec:discussion}

The results extend the connection between semantic organisation,
burstiness and long-range correlations from individual lexical items to
distributed observables in a continuous semantic space. Word burstiness
has long been associated with topicality and heterogeneous word-use
rates \citep{ChurchGale1995,Carpena2009,AltmannPierrehumbertMotter2009,
MontemurroZanette2010,Altmann2012}. Here an event is not the recurrence
of a particular word but a visit to the upper tail of a projection of
the embedding trajectory along a text-specific principal direction.
Semantically related words can consequently contribute to the same
event sequence without recurring as identical lexical items. This
provides a distributed extension of the word-level burstiness
construction of \citet{Altmann2012}. The geometric viewpoint also
connects our analysis to \citet{GromovDang2023}, who studied embedding-based semantic and sentiment trajectories of literary works in twelve
languages using, among other measures, Lyapunov exponents and
entropy--complexity characteristics. Whereas their analysis addresses
dynamical properties and cross-language differences, our focus is on
the temporal recurrence of visits to particular semantic directions.
The null-model comparisons additionally establish which aspects of
that recurrence cannot be explained by static embedding geometry or a
matched rank-sequence scaling exponent.

The relation between semantic variance and burstiness gives a second
perspective on this structure. PCA orders directions by variance, not
by temporal clustering, and the two component spectra are therefore
not interchangeable. Burstiness is generally stronger among the
leading directions but is not monotone in PCA rank, and its maximum
need not occur at PC1. The extension to $K=50$ in the four
representative texts shows that burstiness persists beyond the first
twenty components, with a declining average magnitude. Rather than a
sharply bounded low-dimensional localisation, the data thus suggest a
graded distribution of temporal organisation across semantic
directions. Retaining $K=20$ is primarily a choice of interpretability,
not an empirical boundary beyond which the effect disappears.

The two surrogate models make the separation between static geometry
and temporal organisation particularly explicit. Both surrogates are
permutations of the original token multiset and hence share exactly its
embedding covariance, PCA eigensystem and projected marginals. The
word-shuffled null randomises the ordering, whereas the FGN surrogate
imposes rank-level correlations and is tuned to match the DFA exponent
of the original unfiltered rank sequence. Nevertheless, both controls
remain close to the independent-event baseline for component-level
burstiness. The corpus-level projection ACFs show a corresponding
contrast, and the projection-level DFA analysis of \emph{Great
Expectations} separates the original components from both controls.
Matching the global rank-sequence exponent is therefore insufficient
to reproduce the observed semantic-level temporal organisation. The
conclusion does not depend on the stage at which the exponent is
matched: a surrogate tuned on the filtered token sequence, which then
carries the same rank-level memory as the retained words of the
original, recovers only about $2\%$ of the excess burstiness of the
original texts, mostly along PC1 and PC3
(Section~\ref{sec:corpus_diffs}). Conversely, in the texts in which the
standard surrogate retains more rank-level memory after filtering than
the original, it still shows no excess burstiness.

A useful comparison is provided by \citet{HernandezPerez2024}, who
analysed sentence-level sentiment trajectories obtained with VADER for
50 literary texts. They reported a mean DFA exponent of approximately
$0.62$ for the original sentiment sequences, compared with $0.50$
after shuffling the sentences. This is qualitatively consistent with
our observation of persistent correlations in original semantic
projections and their loss under word shuffling. The observables and
scales, however, differ: sentence-level sentiment is a prescribed
scalar attribute, whereas our token-level PCA projections resolve
multiple text-specific semantic contrasts. Moreover, the FGN control
shows that the semantic persistence measured here is not recovered
simply by reproducing rank-level long-range scaling. The comparison
supports the broader relevance of temporal dependence in semantic and
affective representations, without implying that these observables
capture the same underlying process.

The observed clustering is consistent with a topical interpretation.
Lexical cohesion and the uneven recurrence of content words have long
provided evidence of subtopic structure
\citep{Hearst1997,MontemurroZanette2010,Carpena2009}. In our texts,
semantic poles associated with military discourse in \emph{War and
Peace}, biogeography in \emph{On the Origin of Species}, and other
lexical fields are revisited preferentially within restricted textual
regions. A related emphasis on the arrangement rather than the mere
presence of themes appears in \citet{ScalviniMashaghi2025}, who use
semantic topology to characterise serial, parallel and interleaved
thematic relationships in long-form news articles. Their topological
descriptors and our inter-event-time distributions measure different
properties, but both approaches indicate why temporal arrangement is
an essential part of a description of discourse. In our case, the FGN
surrogate shows specifically that correlations imposed between word
frequency ranks do not suffice to arrange semantically related words
into the same textual regions. Topical organisation is thus a
plausible explanation of the excess burstiness, although the present
controls do not establish it as a unique generative mechanism.

The corpus analysis extends the result beyond the four representative
works. At $q=0.95$ the original exceeds both controls in all 560
text--component comparisons, and the ordering persists over a broad
range of thresholds. The magnitude varies across texts: the eight
non-fiction works in this corpus have a larger mean
original-to-shuffled burstiness ratio than the twenty novels. This
might reflect the more explicit division of some expository works
into relatively specialised subjects, but it remains exploratory.
The texts were selected for variety rather than sampled from balanced
genre populations, so the observed difference is not a general
estimate of a genre effect.

Several limitations qualify these interpretations. Static embeddings
assign one vector to every occurrence of a word, which makes the
permutation controls exact but excludes context-dependent changes of
meaning. Contextual word representations \citep{Devlin2019} can differ
substantially between occurrences \citep{Ethayarajh2019}; applying our method to
them could reveal further semantic dynamics, but shuffling the text
would then alter the embeddings as well and require a different null
construction. Moreover, individual PCA directions can be unstable
when neighbouring eigenvalues are close. The supplementary axis-stability analysis shows that small changes in preprocessing can rotate
such axes and appreciably change an individual $B_k$, even where the
broader pattern remains stable. Exact component maxima should
therefore not be overinterpreted. Rotation-invariant measures over
nearly degenerate PCA subspaces would be a natural extension.

Future work could test this picture on deliberately balanced and
multilingual corpora, extend projection-level DFA beyond the present
representative example, and develop dynamical models of evolving
semantic context that predict both projection correlations and
component-resolved recurrence statistics. In particular, comparing
semantic-topological descriptors with projection burstiness on the
same texts would help establish whether they capture complementary
features of thematic organisation.

In summary, semantic burstiness is distributed over many principal
directions, typically strongest among the earlier components, and
persists with diminishing magnitude beyond the first twenty. Its
separation from both exact-geometry surrogates, together with the
ACF and DFA results, shows that lexical composition and matched
global rank-level scaling do not suffice to reproduce the temporal
structure of semantic projections. The findings support an account
in which ordered recurrence of semantically related material plays
an important role in long-range textual organisation.

\section*{CRediT authorship contribution statement}
\textbf{Mirko Degli Esposti:} Conceptualization, Methodology, Software,
Formal analysis, Writing -- original draft, Writing -- review \& editing.
\textbf{Marcelo A.\ Montemurro:} Conceptualization, Methodology,
Formal analysis, Writing -- review \& editing.
 
\section*{Declaration of competing interest}
The authors declare that they have no known competing financial
interests or personal relationships that could have appeared to
influence the work reported in this paper.
 
\section*{Funding}
This research did not receive any specific grant from funding agencies
in the public, commercial, or not-for-profit sectors.

\section*{Data availability}
All texts are public-domain works from Project Gutenberg. The corpus,
the analysis code, the result tables and the scripts that regenerate
every figure in this paper are available at
\url{https://github.com/mirko-degli-esposti/semantic-burstiness-pca}
and archived on Zenodo, \url{https://doi.org/10.5281/zenodo.23540331}.
Pre-trained GloVe vectors are distributed by Stanford NLP
\citep{Pennington2014}.

\section*{Declaration of generative AI and AI-assisted technologies in the manuscript preparation process}
During the preparation of this work the authors used Claude
(Anthropic) and ChatGPT (OpenAI) to assist with writing and
refactoring the analysis code, to draft and revise passages of the
manuscript, to assist with the literature search, and to check
quantitative statements in the text against the project's own logs
and source documents. After using these tools, the authors reviewed
and edited the content as needed and take full responsibility for the
content of the published article. Every reference was verified
against the original source, and every quantitative result reported
in this paper is produced by the deposited code.

\bibliographystyle{elsarticle-num}

\clearpage
\setcounter{section}{0}\setcounter{figure}{0}\setcounter{table}{0}\setcounter{equation}{0}
\renewcommand{\thesection}{S\arabic{section}}\renewcommand{\theHsection}{S\arabic{section}}
\renewcommand{\thefigure}{S\arabic{figure}}\renewcommand{\theHfigure}{S\arabic{figure}}
\renewcommand{\thetable}{S\arabic{table}}\renewcommand{\theHtable}{S\arabic{table}}
\renewcommand{\theequation}{S\arabic{equation}}\renewcommand{\theHequation}{S\arabic{equation}}
\begin{center}
{\Large\bfseries Supplementary Material}
\end{center}
\section{Inter-event-time distributions for the remaining representative texts}
\label{app:pertext}

Inter-event-time distributions for the three representative texts
other than \emph{War and Peace}, whose corresponding figure appears in
the main text. Equivalent figures for every text of the corpus, at any
threshold quantile, can be regenerated with \texttt{scripts/run\_text.py}
in the code repository.

\begin{figure}[H]
  \centering
  \includegraphics[width=\textwidth]{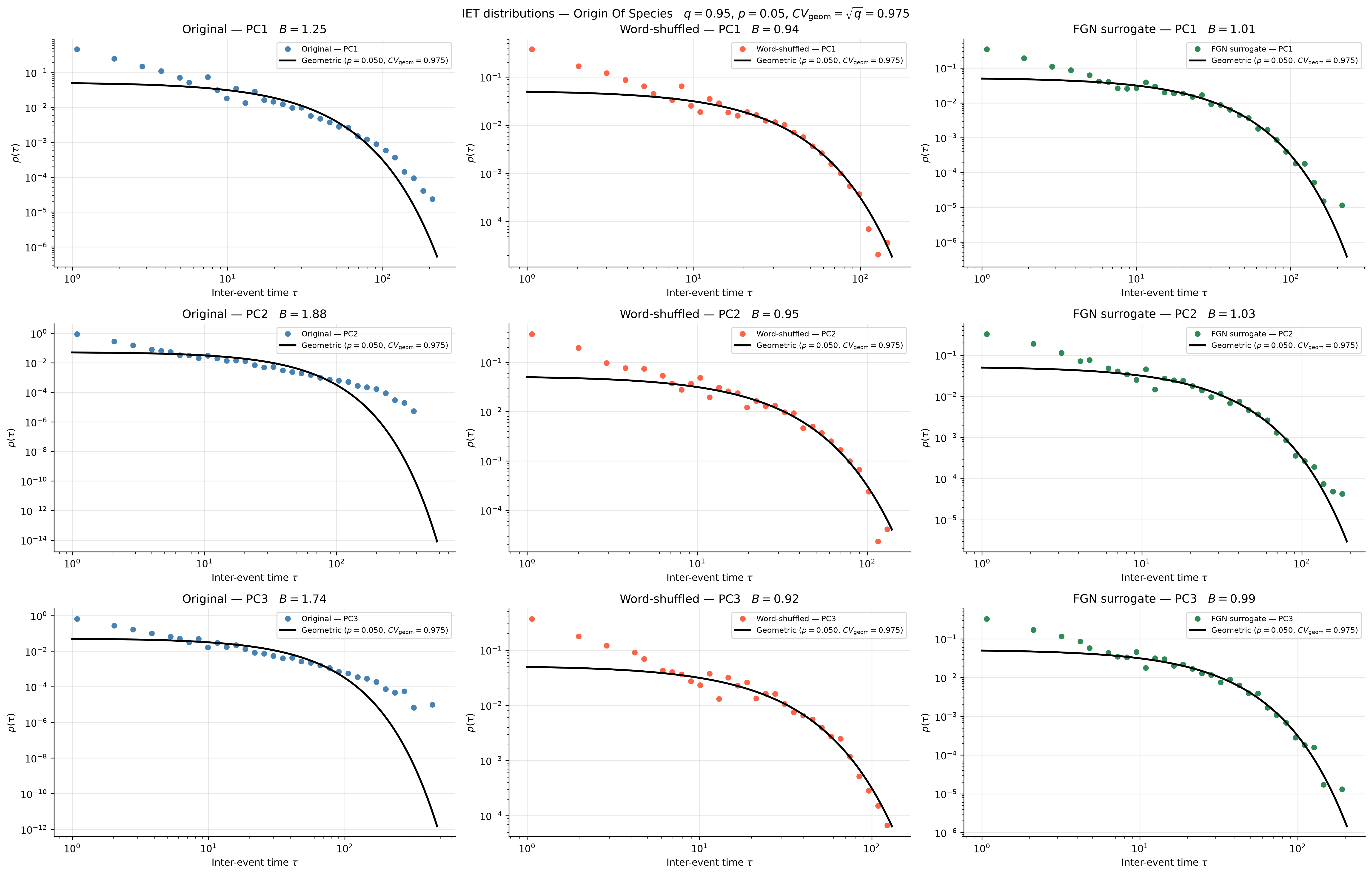}
  \caption{Inter-event time distributions for PC1, PC2, PC3 of
  \emph{On the Origin of Species} ($q=0.95$).
  Same conventions as the corresponding figure in the main text.}
  \label{fig:darwin_iet}
\end{figure}

\begin{figure}[H]
  \centering
  \includegraphics[width=\textwidth]{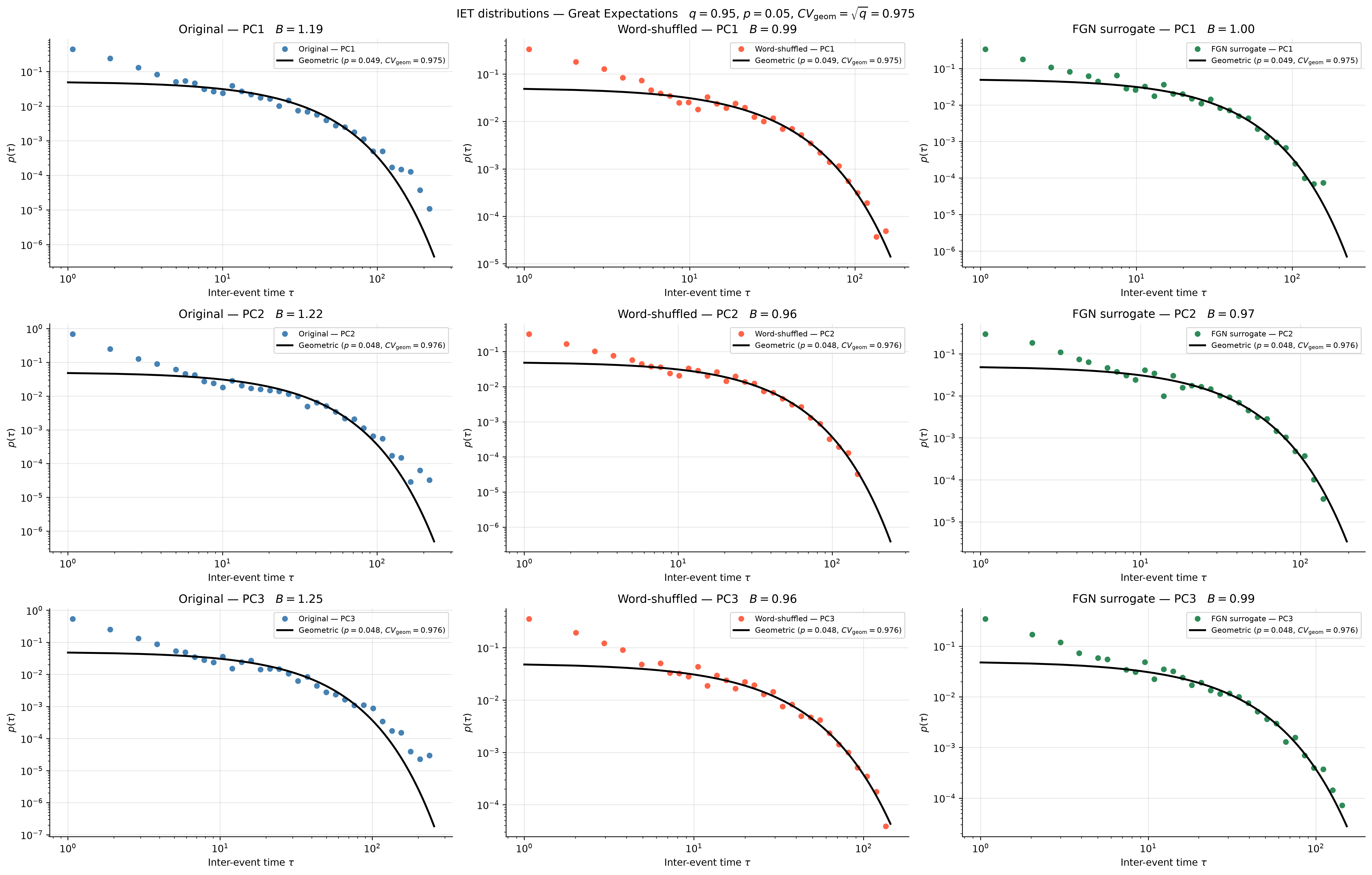}
  \caption{Inter-event time distributions for PC1, PC2, PC3
  of \emph{Great Expectations} ($q=0.95$).
  Same conventions as the corresponding figure in the main text.}
  \label{fig:greatexp_iet}
\end{figure}

\begin{figure}[H]
  \centering
  \includegraphics[width=\textwidth]{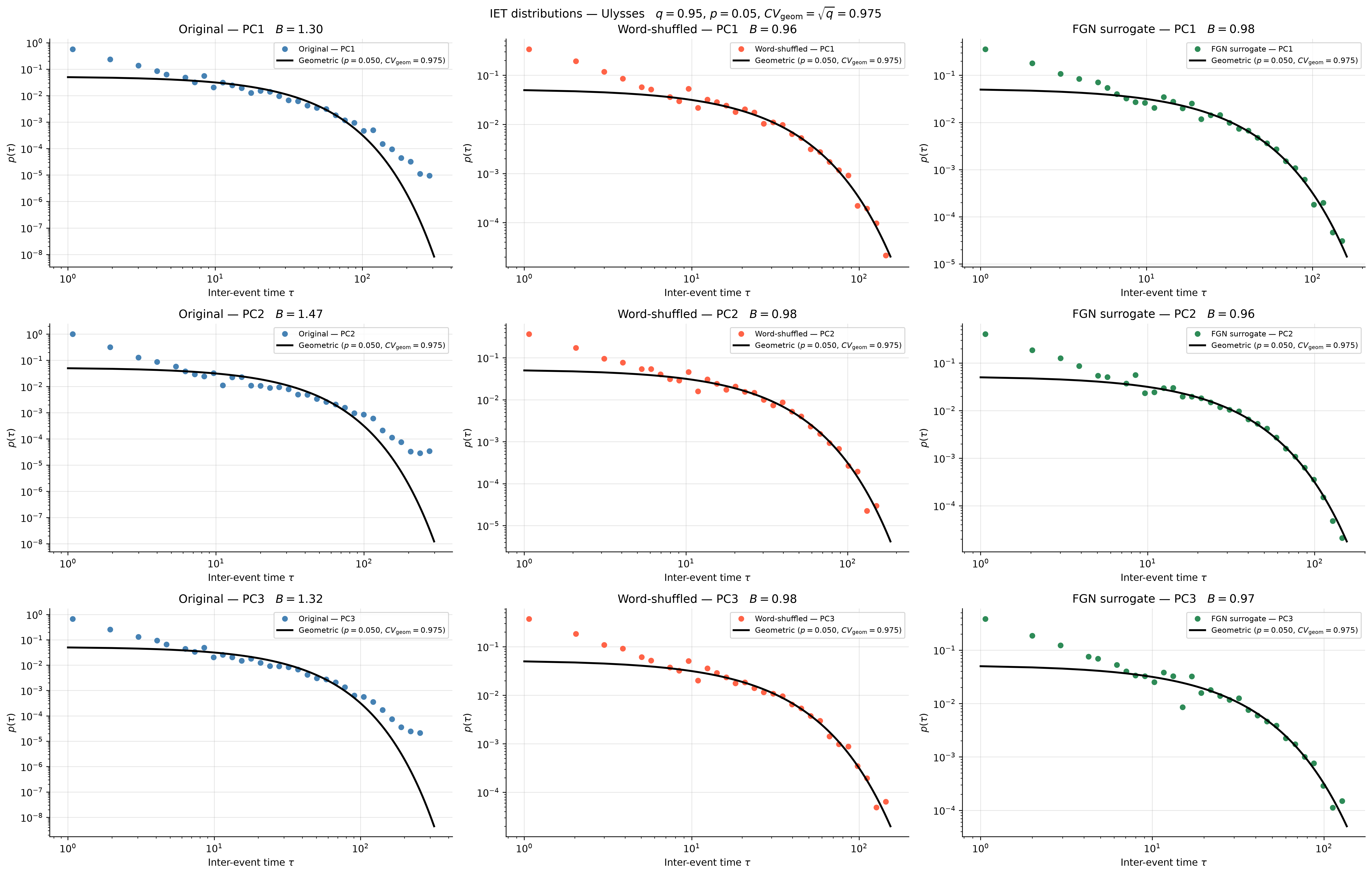}
  \caption{Inter-event time distributions for PC1, PC2, PC3
  of \emph{Ulysses} ($q=0.95$).
 Same conventions as the corresponding figure in the main text.}
  \label{fig:ulysses_iet}
\end{figure}

\clearpage
\section{Additional robustness figures}

\begin{figure}[H]
  \centering
  \includegraphics[width=\textwidth]{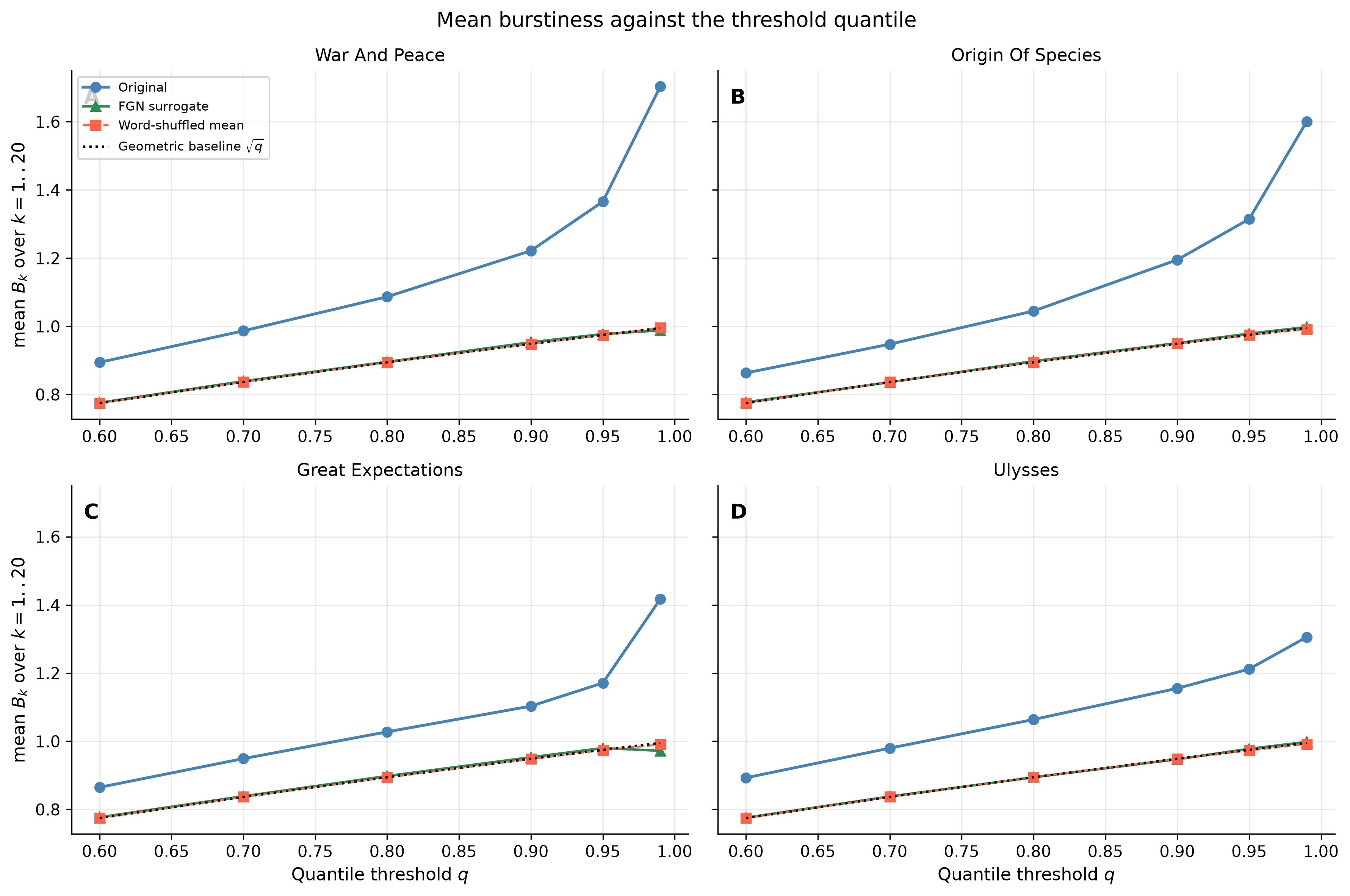}
  \caption{Mean burstiness over $k=1,\ldots,20$ as a function of the
  threshold quantile $q$, for the four representative texts:
  \textbf{A} \emph{War and Peace}, \textbf{B} \emph{On the Origin of
  Species}, \textbf{C} \emph{Great Expectations}, \textbf{D}
  \emph{Ulysses}. The original stays above both controls and above the
  geometric baseline $\sqrt{q}$ throughout $q \in [0.60, 0.99]$, and the
  full ordering $B_k^{\rm orig} > B_k^{\rm FGN}$ and
  $B_k^{\rm orig} > B_k^{\rm surr}$ holds for all 20 components at every
  displayed $q$ in each text. Absolute burstiness grows towards more
  extreme thresholds, as rare visits to the upper tail accentuate
  episode-to-episode clustering.}
  \label{fig:sweep_panel}
\end{figure}

\begin{figure}[H]
  \centering
  \includegraphics[width=\textwidth]{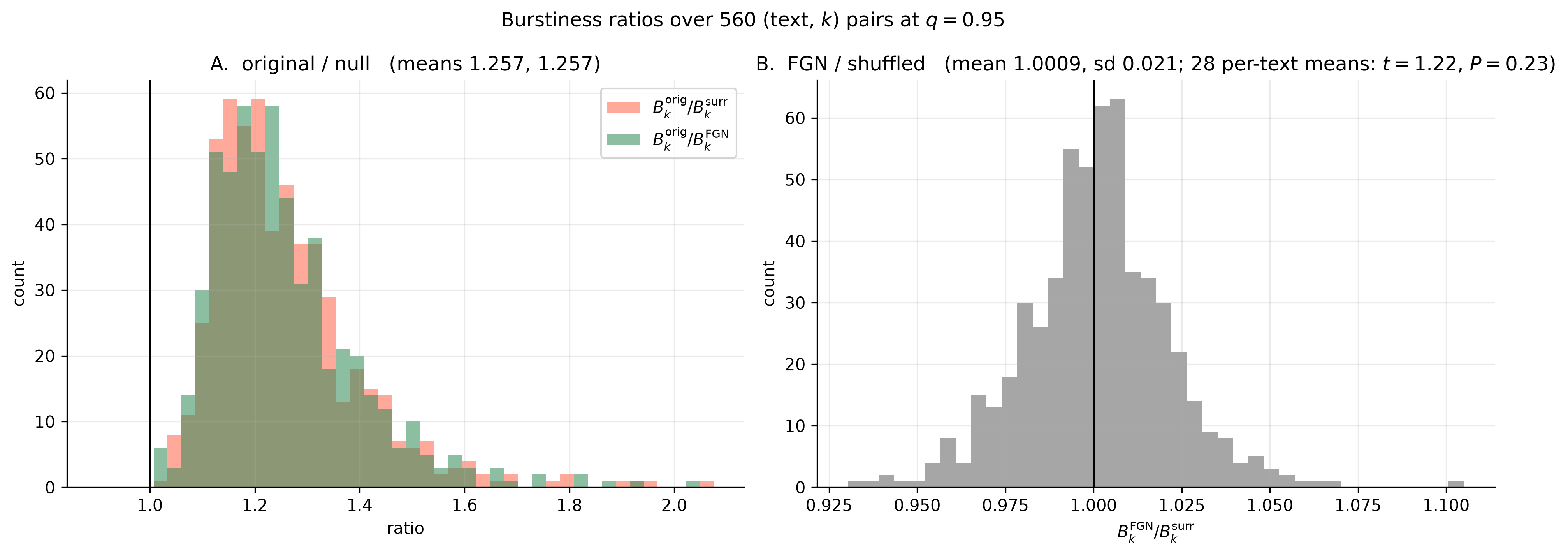}
  \caption{%
    \textbf{Distribution of burstiness ratios over
    560 (text, $k$) pairs at $q = 0.95$.}
    \textbf{Panel A}: distributions of $B_k^{\rm orig}/B_k^{\rm surr}$
    (red) and $B_k^{\rm orig}/B_k^{\rm FGN}$ (green), both
    concentrated well above 1.
    The two distributions are nearly identical, confirming that the
    two null models are equivalently far below the original.
    \textbf{Panel B}: distribution of $B_k^{\rm FGN}/B_k^{\rm surr}$,
    tightly centred on 1 (mean over pairs $1.0009$, sd $0.021$; one-sample
    $t$-test on the 28 per-text means $t = 1.22$, $P = 0.23$), statistically
    indistinguishable from unity, confirming
    $B_k^{\rm FGN} \approx B_k^{\rm surr}$ across the entire corpus.
  }
  \label{fig:corpus_diffs}
\end{figure}

\begin{figure}[H]
  \centering
  \includegraphics[width=0.72\textwidth]{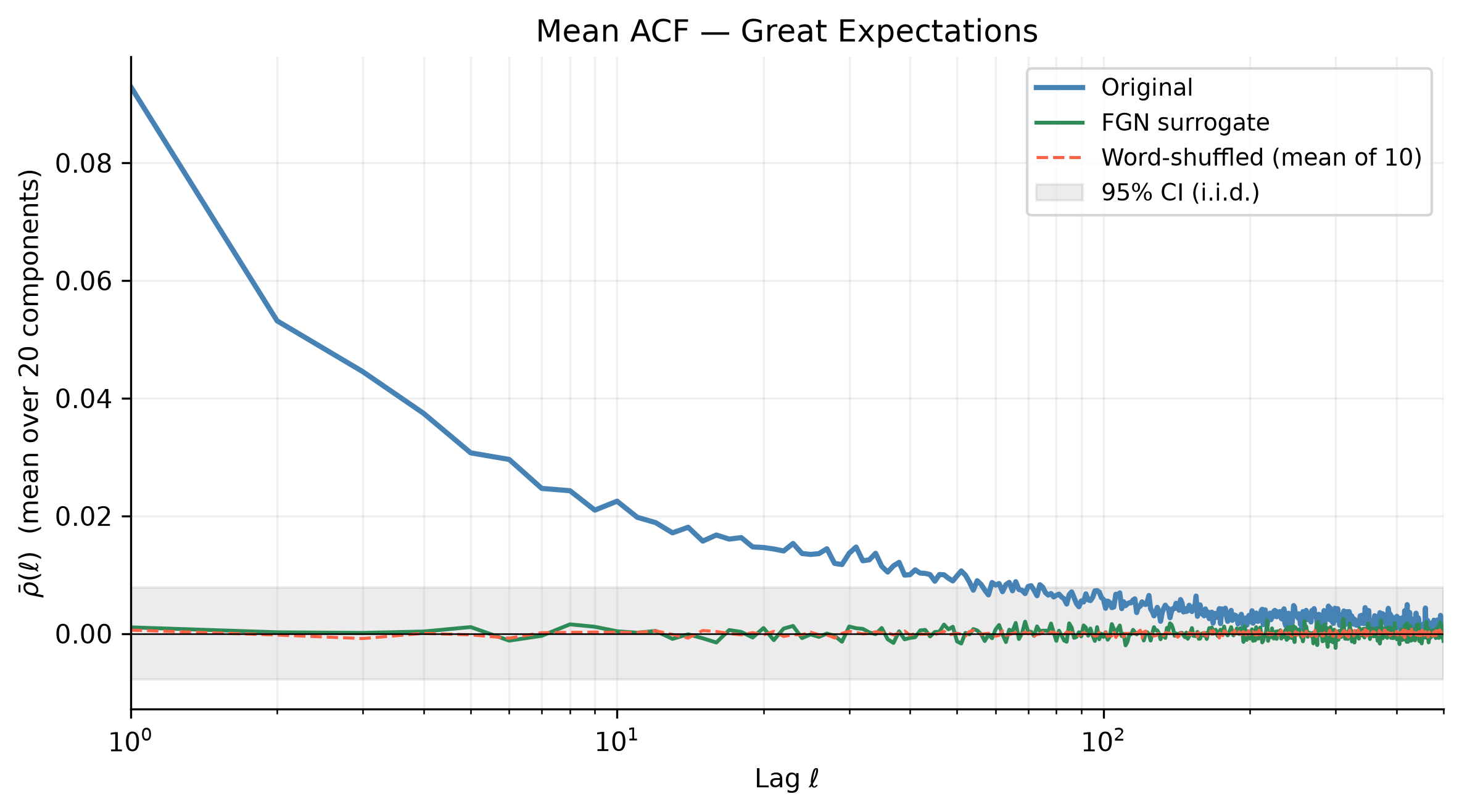}
  \caption{%
    \textbf{Mean ACF $\bar\rho(\ell)$ averaged over all $K=20$ PCA
    components --- \emph{Great Expectations} (Dickens).}
    Same colour conventions as the per-component ACF figure in the main text.
    The original text maintains positive mean autocorrelation well
    beyond the i.i.d.\ band across the full range of lags shown.
    The FGN surrogate and word-shuffled null are
    indistinguishable from the i.i.d.\ reference at all lags,
    confirming that rank-level long-range memory
    (present in the FGN by construction) does not propagate to the
    semantic projection space.
  }
  \label{fig:acf_mean}
\end{figure}

\begin{figure}[H]
  \centering
  \includegraphics[width=\textwidth]{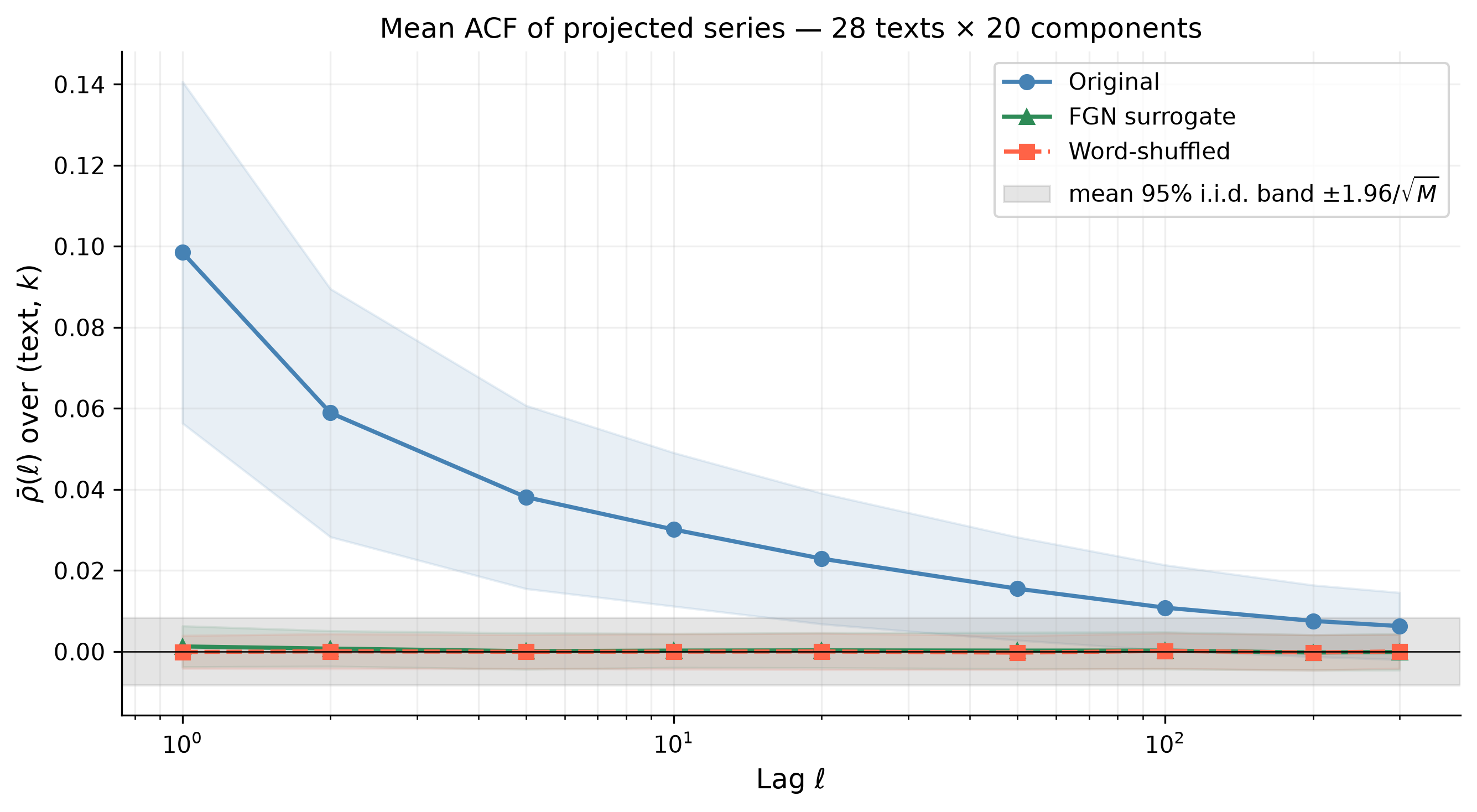}
  \caption{%
    \textbf{Mean ACF $\bar\rho(\ell)$ of projected series across the
    corpus of 28 texts and $K=20$ components, as a function of lag
    $\ell$ (log scale).}
    Shaded bands: $\pm 1$ standard deviation over
    $({\rm text}, k)$ pairs.
    Grey band: mean 95\% i.i.d.\ confidence band
    $\pm 1.96/\sqrt{M}$.
    Blue circles: original filtered text.
    Green triangles: FGN surrogate.
    Red squares: word-shuffled null.
    The original maintains significant positive ACF across all lags
    tested ($\ell = 1, \ldots, 300$).
    Both null models are indistinguishable from zero at all lags.
  }
  \label{fig:acf_corpus_decay}
\end{figure}

\begin{figure}[H]
  \centering
  \includegraphics[width=0.72\textwidth]{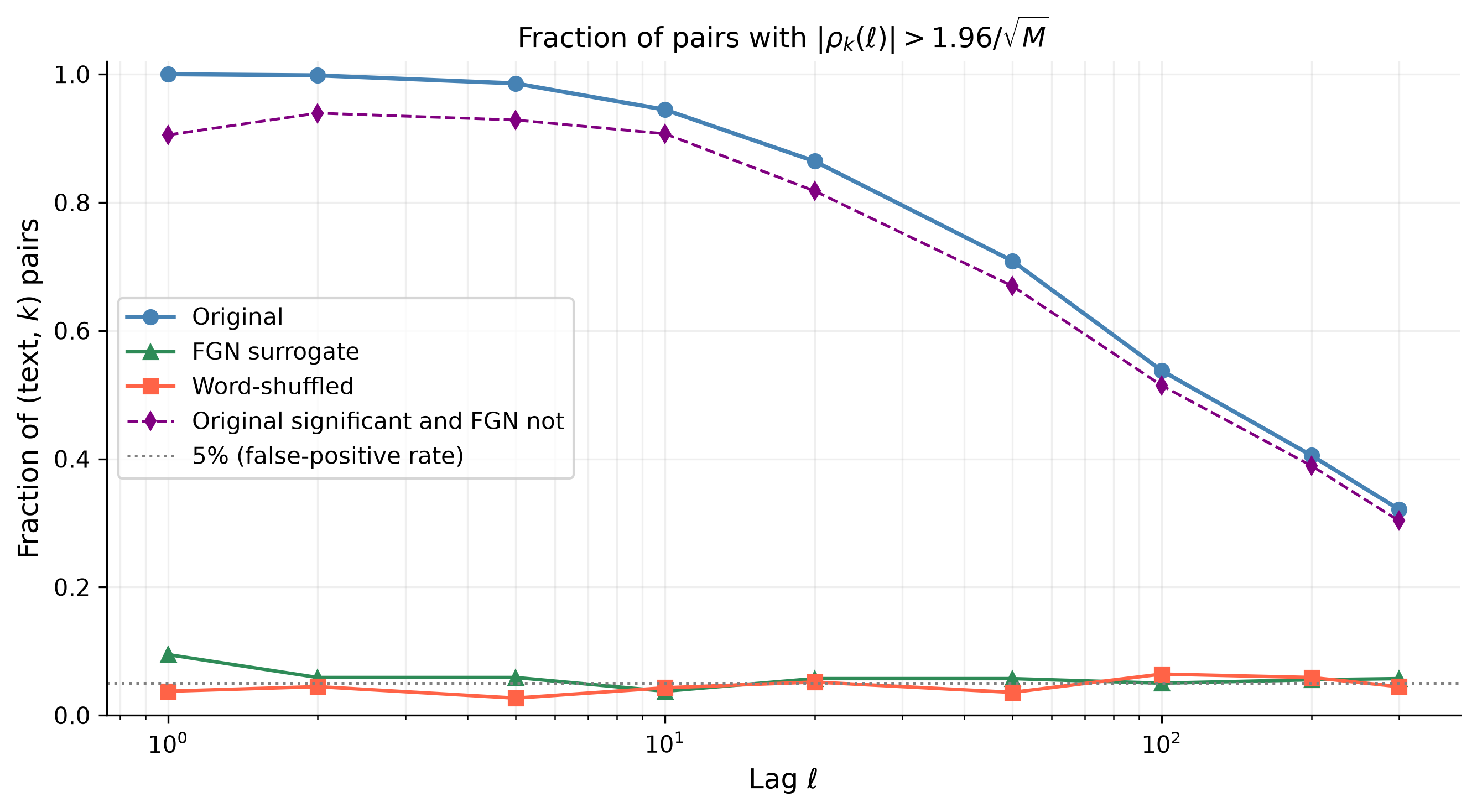}
  \caption{Fraction of text--component pairs with significant ACF as
  a function of lag $\ell$ (log scale). Blue: original; green: FGN;
  red: word-shuffled; purple dashed: original significant and FGN not.
  The original remains clearly separated from both controls through
  lags of order $10^2$. The short-lag FGN excess is concentrated
  disproportionately in PC1, the component most affected by the
  embedding-geometry artefact discussed in the preprocessing section of the main text.}
  \label{fig:acf_corpus_frac}
\end{figure}

\begin{figure}[H]
  \centering
  \includegraphics[width=\textwidth]{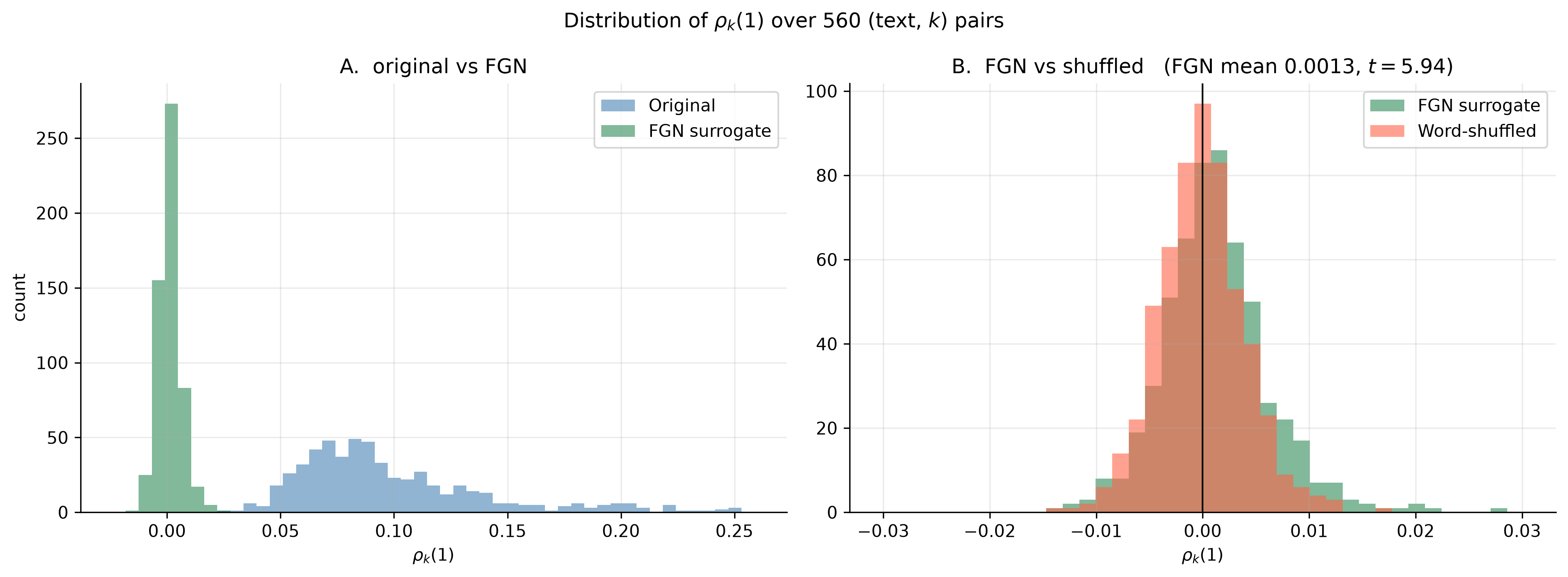}
  \caption{%
    \textbf{Distribution of $\rho_k(1)$ over $560$ (text, $k$) pairs.}
    \textbf{Panel A}: original (blue) and FGN (green).
    The two distributions are completely separated: every original pair
    exceeds its own text-specific i.i.d.\ band (which ranges from
    $0.004$ to $0.013$ across the corpus), while nearly all FGN pairs
    fall inside it.
    \textbf{Panel B}: FGN (green) and word-shuffled (red) on an
    expanded scale.
    Both are centred near zero; the mean FGN value ($0.001$) is
    negligible relative to the orig--null gap ($0.097$).
  }
  \label{fig:acf_corpus_rho1}
\end{figure}
\clearpage
\section{Stability of single-component burstiness}
\label{app:bk_stability}

The burstiness of an individual component depends on the orientation of
its axis, and a principal axis is well determined only when its
eigenvalue is well separated from its neighbours. We tested this on two
preparations of \emph{On the Origin of Species} that differ only in the
treatment of punctuation, apostrophes and editorial material (about
$1\%$ of the tokens). The leading axes are practically identical
($|\cos| \geq 0.997$ between corresponding components for PC1--PC3), and
so are their burstiness values ($B_2 = 1.888$ and $1.848$). PC6, whose
eigenvalue $\lambda_6 = 0.600$ lies between $\lambda_5 = 0.680$ and
$\lambda_7 = 0.564$, rotates by $8.5^\circ$ ($|\cos| = 0.989$), and its
burstiness falls from $1.90$ to $1.25$.

Projecting each preparation onto the axes of the other separates the
two effects. On the same axis the two preparations give nearly the same
value ($B_6 = 1.900$ and $1.873$ on the axis of the first,
$1.270$ and $1.254$ on the axis of the second), whereas the same text
gives $1.90$ or $1.27$ depending on the axis. The change is therefore
due to the rotation of the axis and not to the textual differences.
Because events are defined by the upper $5\%$ tail of the projection,
$B_k$ responds to the relatively few word types with the largest
projections, and a small rotation can move a topically concentrated
group of words (here, the vocabulary of climate and glacial periods) in
or out of the tail.

Values of $B_k$ for single components, and the location of the maximum
over $k$, are therefore interpreted in the main text only for components
that are well separated in the eigenvalue spectrum. The corpus-level
quantities average over components and texts and are stable: over the
13 texts analysed in both preparations, the per-text mean ratio
$\langle B_k^{\rm orig}/B_k^{\rm surr}\rangle_k$ changes by at most
$0.03$. The script that reproduces this test is in the code repository
(\texttt{note/pg\_comparison/origin\_axes.py}).

\clearpage
\section{FGN surrogate matched after filtering}
\label{app:filtered_fgn}

The FGN surrogate of the main text is tuned so that the DFA exponent of
its rank sequence matches that of the original on the unfiltered
sequence of $N$ tokens, and is then filtered. Table~\ref{tab:filtered_fgn}
reports, for every text, the rank-sequence DFA exponent measured with the
estimator of the matching step (order-1 DFA, ten log-spaced scales from
$100$ to $\min(10^4, n/4)$) on the unfiltered sequence and on the $M$
retained tokens, for the original, for the standard surrogate (FGN) and
for a surrogate generated and matched directly on the $M$ retained
tokens (FGN$_M$; bisection range for $\alpha_0$ widened to
$[0.55, 1.0]$). The same rank table, that of the original retained
tokens, is used for all three retained sequences; a constant shift of
the ranks leaves the DFA exponent unchanged. As a check of the estimator,
one random permutation of the retained original tokens gives
$\alpha = 0.489 \pm 0.023$ over the corpus.

Before filtering the standard surrogate matches the original to within
$0.01$ in every text (mean difference $0.000$, sd $0.006$). After
filtering its exponent is lower in 22 of the 28 texts (mean difference
$-0.040$, range $[-0.159, +0.091]$; paired $t = -3.7$, $P < 10^{-3}$);
the original loses $0.063$ on average, the surrogate $0.103$. FGN$_M$
reproduces the exponent of the original retained tokens to within
$0.009$. For \emph{Oliver Twist} the bisection did not converge with the
default seed and was restarted with the next one.

The last three columns give the per-text mean over $k$ of the burstiness
ratio at $q=0.95$. FGN$_M$ has a mean ratio of $1.0062 \pm 0.0012$
(range $[0.989, 1.020]$; one-sample $t = 5.3$ against unity), against
$1.0009 \pm 0.0007$ for the standard surrogate and $1.257 \pm 0.014$ for
the originals. Its excess is concentrated on PC1 and PC3 (mean ratio
$1.022$ on both) and is $1.0045$ on average for $k \geq 4$. In the
projection ACF, $19\%$ of the FGN$_M$ pairs are significant at lag one
($9.5\%$ for the standard surrogate, $100\%$ for the originals), almost
all of them on PC1 and PC3; at lag $100$ the fraction is $6.4\%$, at the
false-positive level. The mean projection DFA exponent is $0.510$
(standard surrogate $0.504$, originals $0.663$). The scripts are
\texttt{scripts/filtered\_rank\_dfa.py} and \texttt{scripts/fgn\_filtered.py}.

\begin{table}[H]
\centering
\caption{Rank-sequence DFA exponents and burstiness ratios of the
original, the standard FGN surrogate and the FGN surrogate matched after
filtering (FGN$_M$). $\alpha_N$: unfiltered original sequence (the
matching target of the standard surrogate); $\alpha_M$: the $M$ retained
tokens; $\alpha_0$: generator exponent of FGN$_M$. Ratios:
$\langle B_k/B_k^{\rm surr}\rangle_k$ at $q = 0.95$. Novels above the
line, non-fiction below. $^{\dagger}$Bisection restarted with the next
seed.}
\label{tab:filtered_fgn}
\scriptsize
\setlength{\tabcolsep}{2.5pt}
\begin{tabular}{l c ccc c ccc}
\toprule
 & & \multicolumn{3}{c}{$\alpha_M$} & & \multicolumn{3}{c}{$\langle B_k/B_k^{\rm surr}\rangle$} \\
\cmidrule{3-5} \cmidrule{7-9}
Text & $\alpha_N$ & orig & FGN & FGN$_M$ & $\alpha_0$ & orig & FGN & FGN$_M$ \\
\midrule
Pride and Prejudice              & 0.642 & 0.640 & 0.553 & 0.637 & 0.59 & 1.150 & 0.989 & 1.003 \\
Jane Eyre                        & 0.682 & 0.631 & 0.555 & 0.636 & 0.61 & 1.196 & 1.003 & 1.001 \\
Wuthering Heights                & 0.721 & 0.616 & 0.625 & 0.625 & 0.58 & 1.146 & 0.995 & 1.004 \\
Don Quixote                      & 0.716 & 0.680 & 0.610 & 0.685 & 0.69 & 1.218 & 0.999 & 1.005 \\
David Copperfield                & 0.719 & 0.715 & 0.621 & 0.715 & 0.72 & 1.195 & 0.998 & 1.009 \\
Great Expectations               & 0.703 & 0.616 & 0.607 & 0.620 & 0.59 & 1.202 & 1.006 & 1.007 \\
Oliver Twist$^{\dagger}$         & 0.721 & 0.569 & 0.613 & 0.577 & 0.61 & 1.199 & 1.002 & 1.004 \\
Middlemarch                      & 0.725 & 0.696 & 0.646 & 0.693 & 0.69 & 1.170 & 1.004 & 1.009 \\
Tess of the d'Urbervilles        & 0.685 & 0.605 & 0.564 & 0.614 & 0.58 & 1.192 & 1.001 & 1.012 \\
Ulysses                          & 0.778 & 0.802 & 0.642 & 0.802 & 0.79 & 1.245 & 1.005 & 1.020 \\
The Rainbow                      & 0.715 & 0.653 & 0.617 & 0.650 & 0.62 & 1.290 & 1.002 & 1.004 \\
Moby Dick                        & 0.698 & 0.694 & 0.561 & 0.695 & 0.66 & 1.277 & 1.002 & 0.999 \\
Frankenstein                     & 0.708 & 0.600 & 0.527 & 0.608 & 0.58 & 1.225 & 0.996 & 0.989 \\
The Jungle                       & 0.773 & 0.662 & 0.698 & 0.665 & 0.62 & 1.283 & 1.001 & 1.006 \\
Dracula                          & 0.768 & 0.715 & 0.681 & 0.719 & 0.68 & 1.246 & 1.000 & 1.011 \\
Vanity Fair                      & 0.717 & 0.666 & 0.620 & 0.663 & 0.66 & 1.197 & 0.999 & 1.005 \\
War and Peace                    & 0.724 & 0.728 & 0.640 & 0.733 & 0.75 & 1.403 & 1.004 & 1.008 \\
Huckleberry Finn                 & 0.736 & 0.642 & 0.612 & 0.646 & 0.62 & 1.288 & 1.002 & 1.007 \\
Tom Sawyer                       & 0.792 & 0.656 & 0.645 & 0.651 & 0.61 & 1.230 & 1.000 & 1.012 \\
The Picture of Dorian Gray       & 0.844 & 0.693 & 0.749 & 0.687 & 0.66 & 1.204 & 0.998 & 0.998 \\
\midrule
On the Origin of Species         & 0.772 & 0.775 & 0.684 & 0.766 & 0.72 & 1.350 & 1.005 & 1.014 \\
The Voyage of the Beagle         & 0.646 & 0.574 & 0.568 & 0.581 & 0.55 & 1.344 & 0.998 & 1.002 \\
Psychology: Briefer Course       & 0.695 & 0.653 & 0.604 & 0.647 & 0.61 & 1.293 & 1.006 & 1.006 \\
A System of Logic                & 0.772 & 0.743 & 0.715 & 0.740 & 0.75 & 1.355 & 1.006 & 1.018 \\
The Analysis of Mind             & 0.752 & 0.628 & 0.630 & 0.621 & 0.58 & 1.340 & 1.001 & 1.011 \\
The Theory of Moral Sentiments   & 0.713 & 0.707 & 0.602 & 0.703 & 0.68 & 1.231 & 0.997 & 1.005 \\
Life on the Mississippi          & 0.742 & 0.700 & 0.650 & 0.696 & 0.65 & 1.297 & 1.001 & 1.001 \\
Island Life                      & 0.861 & 0.692 & 0.783 & 0.694 & 0.65 & 1.440 & 1.003 & 1.005 \\
\bottomrule
\end{tabular}
\end{table}

\clearpage
\section{Full numerical tables}
\begin{table}[H]
\centering
\caption{ACF of projections at lags $\ell = 1$ and $\ell = 100$ for
\emph{Great Expectations}. The i.i.d.\ reference band is
$\pm 1.96/\sqrt{M} = \pm 0.0078$. Values clearly outside the band are
in bold; values that round to the band edge are left unmarked.}
\label{tab:acf_summary}
\small
\begin{tabular}{r rr r rr}
\toprule
 & \multicolumn{2}{c}{$\rho_k(1)$} & \phantom{x}
 & \multicolumn{2}{c}{$\rho_k(100)$} \\
\cmidrule{2-3} \cmidrule{5-6}
$k$ & Orig & FGN && Orig & FGN \\
\midrule
 1 & \textbf{0.067} & \textbf{0.009} && \textbf{0.009} & $-$0.000 \\
 2 & \textbf{0.209} &    0.003 && \textbf{0.022} &    0.004 \\
 3 & \textbf{0.142} &    0.008 && \textbf{0.021} &    0.004 \\
 4 & \textbf{0.127} &    0.002 &&    0.008 & $-$0.001 \\
 5 & \textbf{0.081} & \textbf{0.009} &&    0.007 &    0.001 \\
 6 & \textbf{0.118} &    0.003 &&    0.006 & $-$0.001 \\
 7 & \textbf{0.130} &    0.002 &&    0.000 & $-$0.002 \\
 8 & \textbf{0.093} & $-$0.006 &&    0.008 & \textbf{$-$0.010} \\
 9 & \textbf{0.090} & $-$0.004 && \textbf{0.011} & \textbf{$-$0.012} \\
10 & \textbf{0.084} & $-$0.001 && \textbf{0.010} &    0.004 \\
\midrule
 & \multicolumn{5}{l}{\emph{(k=11..20 all have $|\rho^{\rm FGN}(1)| \le 0.007$)}} \\
\bottomrule
\end{tabular}
\end{table}

\begin{table}[H]
\centering
\caption{DFA scaling exponents $\alpha_k$ of projected series
$x_t^{(k)}$ for \emph{Great Expectations} (Dickens, $\alpha=0.703$).
Every original value exceeds the shuffled reference by more than
$2\sigma$ (the spread over the ten shuffled realisations is
$\sigma\simeq0.004$), and every FGN value lies within $0.05$ of $0.5$.}
\label{tab:dfa_proj}
\scriptsize
\setlength{\tabcolsep}{4pt}
\begin{tabular}{r ccc r ccc}
\toprule
$k$ & $\alpha^{\rm orig}$ & $\alpha^{\rm FGN}$ &
    $\alpha^{\rm surr}$ &
$k$ & $\alpha^{\rm orig}$ & $\alpha^{\rm FGN}$ &
    $\alpha^{\rm surr}$ \\
\midrule
 1 & 0.607 & 0.530$\sim$ & 0.505 &
11 & 0.581 & 0.510$\sim$ & 0.502 \\
 2 & 0.683 & 0.497$\sim$ & 0.499 &
12 & 0.621 & 0.490$\sim$ & 0.497 \\
 3 & 0.696 & 0.514$\sim$ & 0.500 &
13 & 0.583 & 0.495$\sim$ & 0.501 \\
 4 & 0.640 & 0.481$\sim$ & 0.500 &
14 & 0.583 & 0.513$\sim$ & 0.492 \\
 5 & 0.642 & 0.500$\sim$ & 0.499 &
15 & 0.605 & 0.494$\sim$ & 0.500 \\
 6 & 0.634 & 0.482$\sim$ & 0.500 &
16 & 0.615 & 0.520$\sim$ & 0.499 \\
 7 & 0.644 & 0.513$\sim$ & 0.497 &
17 & 0.598 & 0.518$\sim$ & 0.488 \\
 8 & 0.634 & 0.483$\sim$ & 0.502 &
18 & 0.631 & 0.476$\sim$ & 0.502 \\
 9 & 0.656 & 0.483$\sim$ & 0.501 &
19 & 0.610 & 0.479$\sim$ & 0.500 \\
10 & 0.640 & 0.517$\sim$ & 0.500 &
20 & 0.591 & 0.500$\sim$ & 0.498 \\
\midrule
\multicolumn{4}{l}{\textbf{Mean}} &
\multicolumn{4}{l}{$\bar\alpha^{\rm orig}=0.625\pm0.032$ \quad
                   $\bar\alpha^{\rm FGN}=0.500\pm0.016$ \quad
                   $\bar\alpha^{\rm surr}=0.499\pm0.004$}\\
\bottomrule
\end{tabular}
\end{table}

\clearpage
\section{Geometric distribution as the inter-event-time null}
\label{app:geometric}

Consider a binary sequence $b_1,b_2,\ldots,b_M$ in which each $b_t$
is independently Bernoulli$(p)$. An event occurs at position $t$ when
$b_t=1$. The inter-event time $\tau$ between two successive events is
the number of positions (including the terminal event) from one event
to the next. The probability of $\tau=k$ requires
$b_{t+1}=\cdots=b_{t+k-1}=0$ and $b_{t+k}=1$:
\begin{equation}
  \Pr(\tau=k)=(1-p)^{k-1}p, \qquad k=1,2,3,\ldots,
  \label{eq:si_geom_dist}
\end{equation}
which is the geometric distribution with parameter $p$. Its mean,
variance and coefficient of variation are
\begin{equation}
  \langle\tau\rangle=\frac{1}{p},
  \qquad
  \mathrm{Var}(\tau)=\frac{1-p}{p^2},
  \qquad
  \mathrm{CV}_{\rm geom}(p)=\sqrt{1-p}.
  \label{eq:si_geom_cv}
\end{equation}
The exponential distribution $p(\tau)\propto e^{-\lambda\tau}$ arises
in the limit $p\to0$, with $\lambda=-\log(1-p)\approx p$, and has
$\mathrm{CV}=1$ (Poisson process). For $p=0.05$,
$\mathrm{CV}_{\rm geom}=0.975\approx1$, so the Poisson approximation is
adequate. For $p=0.5$, $\mathrm{CV}_{\rm geom}=0.707$, and the Poisson
approximation fails badly.

\clearpage
\section{Additional construction and DFA figures}
\label{app:additional_figures}

\begin{figure}[H]
  \centering
  \includegraphics[width=\textwidth]{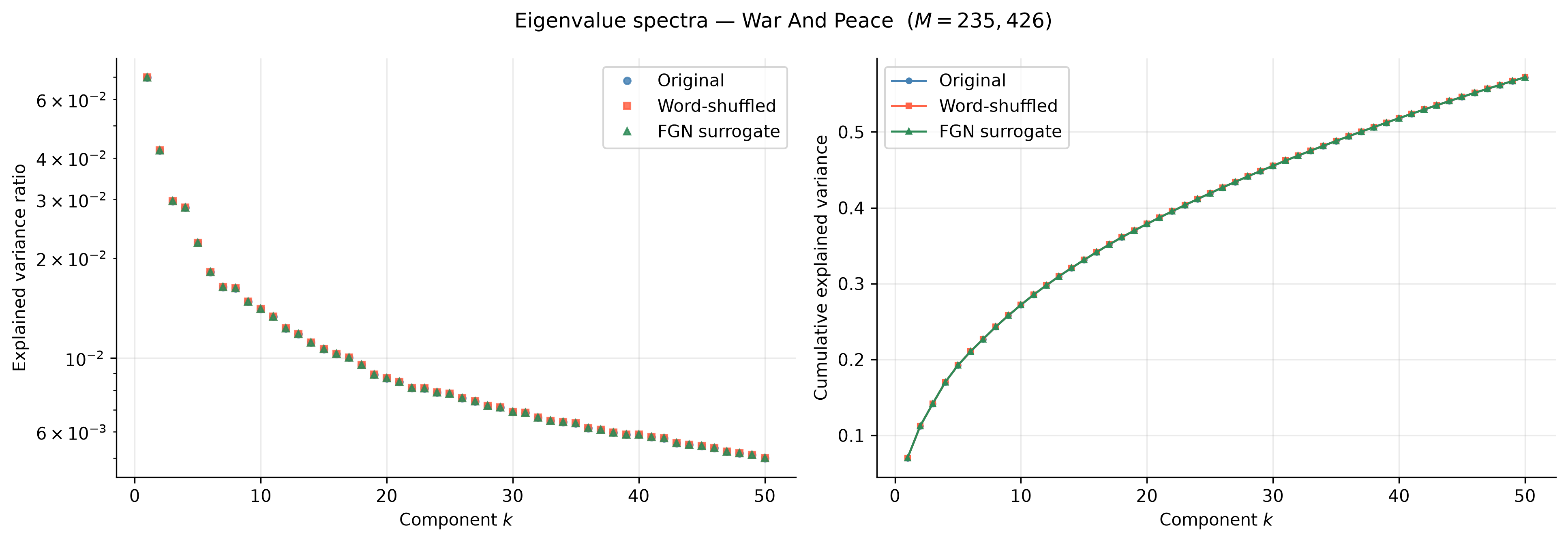}
  \caption{Eigenvalue spectra and cumulative explained variance for
  \emph{War and Peace}. The three curves coincide because both
  surrogates are permutations of the original token sequence, so all
  three share the same covariance matrix and hence the same spectrum.
  The figure therefore verifies the covariance-preserving construction
  rather than reporting a difference between sequences.}
  \label{fig:wrnpc_spectra}
\end{figure}

\begin{figure}[H]
  \centering
  \includegraphics[width=\textwidth]{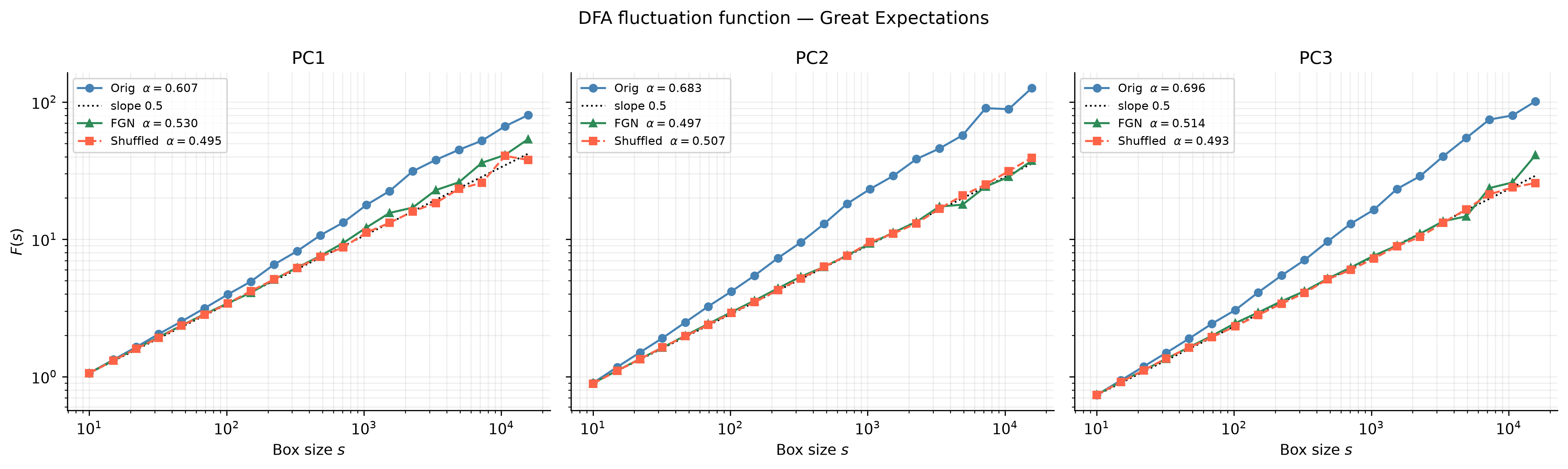}
  \caption{%
    \textbf{Log-log DFA fluctuation function $F(s)$ for PC1, PC2,
    and PC3 --- \emph{Great Expectations} (Dickens).}
    Each panel shows one component; the estimated exponent $\alpha_k$
    is given in the legend. Dotted line: reference slope $0.5$. The
    original text (blue) has a consistently steeper slope than both
    null models, which are parallel to the $\alpha=0.5$ reference
    across all box sizes $s$.
  }
  \label{fig:dfa_loglog}
\end{figure}

\end{document}